\documentclass[aps,rmp,twocolumn,groupedaddress,showpacs]{revtex4}

\usepackage{graphics}
\usepackage{rotating}
\usepackage{verbatim}
\usepackage{url}
\usepackage{hyperref}

\def\aap{Astron.\  Astrophys.\ }
\def\aaps{Astron.\  Astrophys.\  Suppl.\ Ser.\ }
\def\apj{Astroph.\  J.\ }
\def\apjl{Astroph.\  J.\ Lett.\ }

\def\aplett{Astrophysics Lett.\ }

\def\jcp{J. Chem.\ Phys.\ }

\def\pra{Phys.\ Rev.\ A }
\def\prd{Phys.\ Rev.\ D }
\def\prl{Phys.\ Rev.\ Lett.\ }

\def\jms{J. Mol.\ Spectrosc.\ }
\def\jpb{J. Phys.\ B }

\def\lrr{Living\ Rev.\ Relativity }
\def\nat{Nature\ }

\def\molp{Mol.\ Phys.\ }

\def\jctc{J. Chem.\ Theor.\ Comp.\  }

\def\rmp{Rev. Mod.\ Phys. }

\def\mnras{Mon.\ Not.\ Roy.\ Astron.\ Soc.\ }

\def\eujpd{Eur.\ J. \ Phys. \ D }
\def\memsai{Mem.\ Soc.\ Astron.\ Ital.\ }
\newcommand{\wn}{\ensuremath{\mathrm{cm^{-1}}}}

\begin{document}

\title{Search for a drifting proton--electron mass ratio from H$_2$}

\author{W. Ubachs, J. Bagdonaite,}
\affiliation{Department of Physics and Astronomy, LaserLaB, VU University, De Boelelaan 1081, 1081 HV Amsterdam, The Netherlands}

\author{E. J. Salumbides,}
\affiliation{Department of Physics and Astronomy, LaserLaB, VU University, De Boelelaan 1081, 1081 HV Amsterdam, The Netherlands}
\affiliation{Department of Physics, University of San Carlos, Cebu City 6000, Philippines}

\author{M. T. Murphy,}
\affiliation{Centre for Astrophysics and Supercomputing, Swinburne University of Technology, Melbourne, Victoria 3122, Australia}

\author{L. Kaper}
\affiliation{Astronomical Institute Anton Pannekoek, Universiteit van Amsterdam,
Postbus 94249, 1090 GE Amsterdam, The Netherlands}

\date{\today}

\begin{abstract}
An overview is presented of the H$_2$ quasar absorption method to search for a possible variation of the proton--electron mass ratio $\mu=m_p/m_e$ on a cosmological time scale. The method is based on a comparison between wavelengths of absorption lines in the H$_2$ Lyman and Werner bands as observed at high redshift with wavelengths of the same lines measured at zero redshift in the laboratory. For such comparison sensitivity coefficients to a relative variation of $\mu$ are calculated for all individual lines and included in the fitting routine deriving a value for $\Delta\mu/\mu$.  Details of the analysis of astronomical spectra, obtained with large 8--10 m class optical telescopes, equipped with high-resolution echelle grating based spectrographs, are explained. The methods and results of the laboratory molecular spectroscopy of H$_2$, in particular the laser-based metrology studies for the determination of rest wavelengths of the Lyman and Werner band absorption lines, are reviewed.
Theoretical physics scenarios delivering a rationale for a varying $\mu$ will be discussed briefly, as well as alternative spectroscopic approaches to probe variation of $\mu$, other than the H$_2$ method. Also a recent approach to detect a dependence of the proton-to-electron mass ratio on environmental conditions, such as the presence of strong gravitational fields, will be highlighted. Currently some 56  H$_2$ absorption systems are known and listed. Their usefulness to detect $\mu$-variation is discussed, in terms of column densities and  brightness of background quasar sources, along with future observational strategies. The astronomical observations of ten quasar systems analyzed so far set a constraint on a varying proton-electron mass ratio of $|\Delta\mu/\mu| < 5 \times 10^{-6}$ (3-$\sigma$), which is a null result, holding for redshifts in the range $z=2.0-4.2$. This corresponds to look-back times of 10--12.4 billion years into cosmic history.
Attempts to interpret the results from these 10 H$_2$ absorbers in terms of a spatial variation of $\mu$ are currently hampered by the small sample size and their coincidental distribution in a relatively narrow band across the sky.

\end{abstract}


\maketitle

\section{Introduction}
\label{intro}

For a long time scientific thought had been rooted in the belief that the laws of nature are universal and eternal: they are the rules responsible for the evolution of every part of the universe but they are never subject to a change themselves, nor do they depend on a specific location. Physical laws are encrypted with the so-called fundamental constants that are regarded as fundamental because they can only be determined by experiment. There is no theory predicting the values of the fundamental constants, while at the same time their specific values are finely tuned to accommodate a Universe with complexity~\cite{Carr1979}. Testing the immutability of the laws is equivalent to probing space-time variations and local dependencies of the values of the fundamental constants.
Dirac was one of the first who questioned whether the constants are simply mathematical numbers or if they can be decoded and understood within a context of a deeper cosmological theory~\cite{Dirac1937}. As a result of this reasoning he postulated that one of the fundamental constants, the constant describing the strength of gravity, would vary over cosmic time. The specific hypothesis of the scaling relation $G \propto 1/t$ has been extensively discussed, and falsified over time~\cite{Barrow_book}.

Today, given the substantial body of theoretical and experimental work produced over the past decades, the invariance of physical laws hardly classifies as an axiom anymore, but rather is a testable and therewith falsifiable scientific question with exciting implications and vigorous research programs.
The search for varying fundamental constants, or for putting constraints on their variation, has become part of experimental science.
Tests for variations of fundamental constants are focused on dimensionless constants for the searches to be operationally viable~\cite{Uzan2011}.
Important milestones were achieved in the domain of observational astrophysics, such as establishing the alkali-doublet method to probe the constancy of the fine structure constant $\alpha$ using cosmologically-distant objects~\cite{Savedoff1956,Bahcall1967}, followed by the more general and more accurate many-multiplet method~\cite{Dzuba1999}. The latter method is at the basis of the ground-breaking studies on $\alpha$-variation on a cosmological time scale by \citet{Webb1999}, as well as the search for a spatial effect on varying constants~\cite{Webb2011}.

\citet{Thompson1975} suggested that the wavelengths of molecular hydrogen transitions could be used to probe a possible variation of the ratio of the proton--electron inertial masses. Initial studies to probe H$_2$ at high redshift employed telescopes in the 4m class; the Multiple Mirror Telescope (MMT) at Arizona, the Anglo-Australian Telescope (AAT) in Australia, and the Cerro Tololo Inter-American Observatory (CTIO) at La Serena, Chile. The development of the 8--10m class optical telescopes (Keck at Hawaii and ESO-VLT at Paranal, Chile) constituted an instrumental basis required to implement the aforementioned methods in extragalactic studies at high precision.

Although the analysis of Big Bang nuclear synthesis of elements~\cite{Kolb1986,Barrow1987}, of cosmic microwave background radiation patterns~\cite{Landau2010}, of isotopic composition of ores found in the Oklo mine in Gabon~\cite{Shlyakhter1976}, and of elemental abundances in meteorites~\cite{Olive2002} have resulted in constraints on varying constants, spectroscopic methods currently are the preferred testing grounds for probing a possible variation of a fundamental constant. The main reason is that frequency or wavelength measurements can be performed at extreme precision. This holds in particular for advanced laboratory metrology studies with ultrastable lasers, ion traps, laser-cooling methods, frequency comb lasers and atomic fountain clocks as ingredients. A simultaneous measurement of optical transitions in Hg$^+$ and Al$^+$ ions has produced a limit on a drift rate for the fine structure constant in the present epoch of $\dot\alpha/\alpha = (-1.6 \pm 2.3) \times 10^{-17}$~yr$^{-1}$~\cite{Rosenband2008}.
Recently, laboratory constraints on both $\alpha$ and $\mu$ have been determined by measuring two optical transitions in $^{171}$Yb$^+$ ions~\cite{Godun2014,Huntemann2014}. Assuming a linear drift rate, somewhat intertwined constraints of $\dot\mu/\mu < 10^{-16}$~yr$^{-1}$ and $\dot\alpha/\alpha < 10^{-17}$~yr$^{-1}$ were derived.
A direct $\mu$ constraint was derived from a molecular study (measuring a transition in SF$_6$) resulting in $\dot\mu/\mu = (-3.8 \pm 5.6) \times 10^{-14}$~yr$^{-1}$~\cite{Shelkovnikov2008}, which is less constraining but also less model-dependent~\cite{Flambaum2006}.

Astrophysical approaches bear the advantage that very large time intervals are imposed on the measurements of transition wavelengths at high redshift, leading to an enhancement in sensitivity by 10$^9$--10$^{10}$ for probing a rate of change, with respect to pure laboratory searches. This is under the assumption that a fundamental constant would vary linearly in time.
Search for a variation of $\alpha$ is generally investigated via the measurement of atomic lines and variation of $\mu$ through the measurement of molecular lines. As will be discussed, the H$_2$ lines are not the most sensitive probes. Other molecules, in particular the ammonia~\cite{Flambaum2007} and methanol~\cite{Jansen2011} molecules, are more sensitive testing grounds to probe a variation of $\mu$, but H$_2$ has the advantage that it is ubiquitously observed in the universe, and up to high redshift.

The $B^1\Sigma_u^+ - X^1\Sigma_g^+$ Lyman and $C^1\Pi_u - X^1\Sigma_g^+$ Werner band lines are strong dipole-allowed absorption lines of the H$_2$ molecule covering the wavelength window $\lambda = 910 - 1140$ \AA. These wavelengths cannot be observed with ground-based telescopes in view of the opaqueness of the Earth's atmosphere for wavelengths $\lambda < 3000$ \AA. When observing the Lyman and Werner band lines at redshifts $z$, their wavelengths are multiplied by a factor of $(1+z)$ due to the expansion of space on a cosmological scale. Hence, for redshifts $z \gtrsim 2$ the H$_2$ lines are shifted into the atmospheric transparency window so that the lines can be observed from ground-based telescopes, such as the Very Large Telescope (VLT) of the European Southern Observatory (ESO) or the Keck telescope on Mauna Kea at Hawaii. Alternatively, the H$_2$ absorptions can be observed from outside the Earth's atmosphere with the Hubble Space Telescope. In the latter case H$_2$ absorption at redshifts $z<2$ can be detected.

The main focus of the present review is on a variation of the proton--electron mass ratio $\mu$, a dimensionless constant whose value from laboratory experiments is known with a $4\times10^{-10}$ relative precision and listed in the CODATA 2010 release of the recommended values of the fundamental physical constants~\cite{CODATA2010}. A four-fold improvement over the precision of the current recommended value of $\mu$ was reported recently~\cite{Sturm2014}, leading to
\begin{equation}
\label{mu-def}
\mu \equiv \frac{m_p}{m_e} = 1836.152\,673\,77\,(17).
\end{equation}
It should be realized that searches for a varying constant $\mu$ do not necessarily imply a determination of its value; in fact in none of the studies performed so far was this the strategy. Rather, a variation of $\mu$ is probed as a differential effect
\begin{equation}
\label{mu-diff}
 \Delta \mu = \mu_z - \mu_0
\end{equation}
where $\mu_z$ is the value of the proton--electron mass ratio at a certain redshift $z$ and $\mu_0$ is the value in the current epoch, at zero redshift.
The differential effect is only investigated in relative terms, i.e.~values of $\Delta\mu/\mu$ are determined in the studies~\cite{Jansen2014}.

In the present paper an account is given of the high-redshift observations of H$_2$ absorption spectra in the sightline of quasar\footnote{We use the term ``quasar'' in this work interchangeably with the term “quasi-stellar object”, or “QSO”.} sources and the constraints on a varying proton--electron mass ratio that can be derived from these observations when comparing to laboratory spectra of H$_2$. Also a review is given of the laser-spectroscopic investigations of the Lyman and Werner band lines as performed in the laboratory, both for the H$_2$ and HD isotopomers. Limitations and prospects of the astronomical observations and of the H$_2$ method will be discussed, while a status report is presented on the objects suitable for such studies.

\section{Theories of a varying $\mu$}
\label{theory}

A variety of theoretical scenarios exists in which fundamental constants are allowed to vary: unification theories that include extra dimensions and theories embracing fundamental scalar fields.
Theories postulating additional space-time dimensions date back to the Kaluza--Klein concept of unifying field theories of electromagnetism and gravity in 5 dimensions. Modern string theories are formulated in as many as ten dimensions and postulate a Klein compactification~\cite{Klein1926} to comply with the perceivable 4-dimensional universe. In this dimensional reduction, the conventional constants appear as  projected effective parameters that may be easily varied within the context of cosmological evolution of the universe, resulting in variations of the effective constants as measured in 4-dimensional space-time~\cite{Martins2010}. Interestingly,  observation
of varying constants would have important implications in the present debate on string theory landscapes or multiverses~\cite{Schellekens2013}.

In another class of theories, introduced by  \citet{Bekenstein1982}, a cosmological variation in the electric charge $e$ is produced by an additional scalar field, or a dilaton field, which may weakly couple to matter. A more general approach was proposed by Barrow, Sandvik, and Magueijo~\cite{Sandvik2002,Barrow2002a}, who created a self-consistent cosmological model with varying $\alpha$. In this model any substantial $\alpha$ changes are suppressed with the onset of dark energy domination, a concept with far-reaching implications for laboratory searches of varying constants. \citet{Barrow2005} subsequently described a scenario that, by inducing variations in the mass parameter for the electron, specifically addresses variation of $\mu$. In these theories, variation of constants is driven by the varying matter density in the universe. This led to the development of the so-called chameleon scenarios where inhomogeneities in the constants could be observed by probing environments with very low local mass densities, in contrast to the high-density environment of Earth-based experiments~\cite{Khoury2004}. Such scenarios of environmental dependencies of fundamental constants was extended to cases where $\alpha$ and $\mu$ were predicted to depend on local gravitational fields \cite{Flambaum2008}, by definition implying a breakdown of the equivalence principle.

As an alternative to scenarios producing variation of constants due to massless dark-energy-type fields, \citet{Stadnik2015} recently linked the variation to a condensate of massive dark-matter particles. It is stipulated that this causes a slow drift as well as an oscillatory evolution of the constants of nature.

In the above-mentioned theories by Barrow and co-workers~\cite{Sandvik2002,Barrow2002a} it is implied that the values of fundamental constants are connected to the matter density $\Omega_m$ and dark energy density $\Omega_{\Lambda}$ composition of the Universe. The H$_2$ absorption method is typically applied for redshifts $z>2$, i.e.~for a Universe with a matter density of $\Omega_m>0.9$ (including dark matter). In contrast, the laboratory searches for varying $\mu$, such as those of \citet{Godun2014} and \citet{Huntemann2014}, are performed in the present cosmic epoch with $\Omega_m = 0.32$, hence quite different. Even if laboratory studies based on atomic clocks achieve competitive constraints on a rate-of-change of a varying constant, like $\dot\mu/\mu$, this does not imply that the same physics is probed: the cosmological conditions setting the values of the fundamental constants are likely to be very different.

As was pointed out originally by \citet{Born1935} the fine structure constant $\alpha$ and the proton--electron mass ratio $\mu$ are the two dimensionless parameters that describe the gross structure of atomic and molecular systems.
While the fine structure constant $\alpha$ is a measure of a fundamental coupling strength, the dimensionless constant $\mu$ might be considered less fundamental because the mass of a composite particle is involved.
However, since the gluon field that binds quarks inside the proton is responsible for virtually all of its mass, $\mu$ is sensitive to the ratio of the chromodynamic scale with respect to the electroweak scale~\cite{Flambaum2004}. Thus $\mu$ may be viewed as a parameter connecting fundamental coupling strengths of different forces.
Various theoretical scenarios have been described relating the possible variation of both constants, and in most schemes either relying on Grand Unification~\cite{Calmet2002,Langacker2002} or on string theory~\cite{Dent2003}, the rate of change in the proton-to-electron mass ratio $\Delta\mu/\mu$ is found much larger than the rate of change in the fine structure constant $\Delta\alpha/\alpha$. This makes the search for a varying $\mu$ a sensitive testing ground to probe for variations of fundamental constants.

In the context of spectroscopy and spectral lines the proton and electron masses and their ratio should be considered as \emph{inertial} masses, which is how masses enter the Schr\"{o}dinger equation in the calculation of atomic and molecular level structure.  The difference between inertial versus gravitational masses is of relevance when searching for a variation of fundamental constants, since that implies a space-time dependence of the laws of physics, i.e.~a violation of Einstein Equivalence principle, the fundamental concept by which inertial and gravitational masses are considered identical.

\section{The spectrum of H$_2$ as a test ground}
\label{H2-spec}

\begin{figure}
\centering
\includegraphics[width=1.\columnwidth]{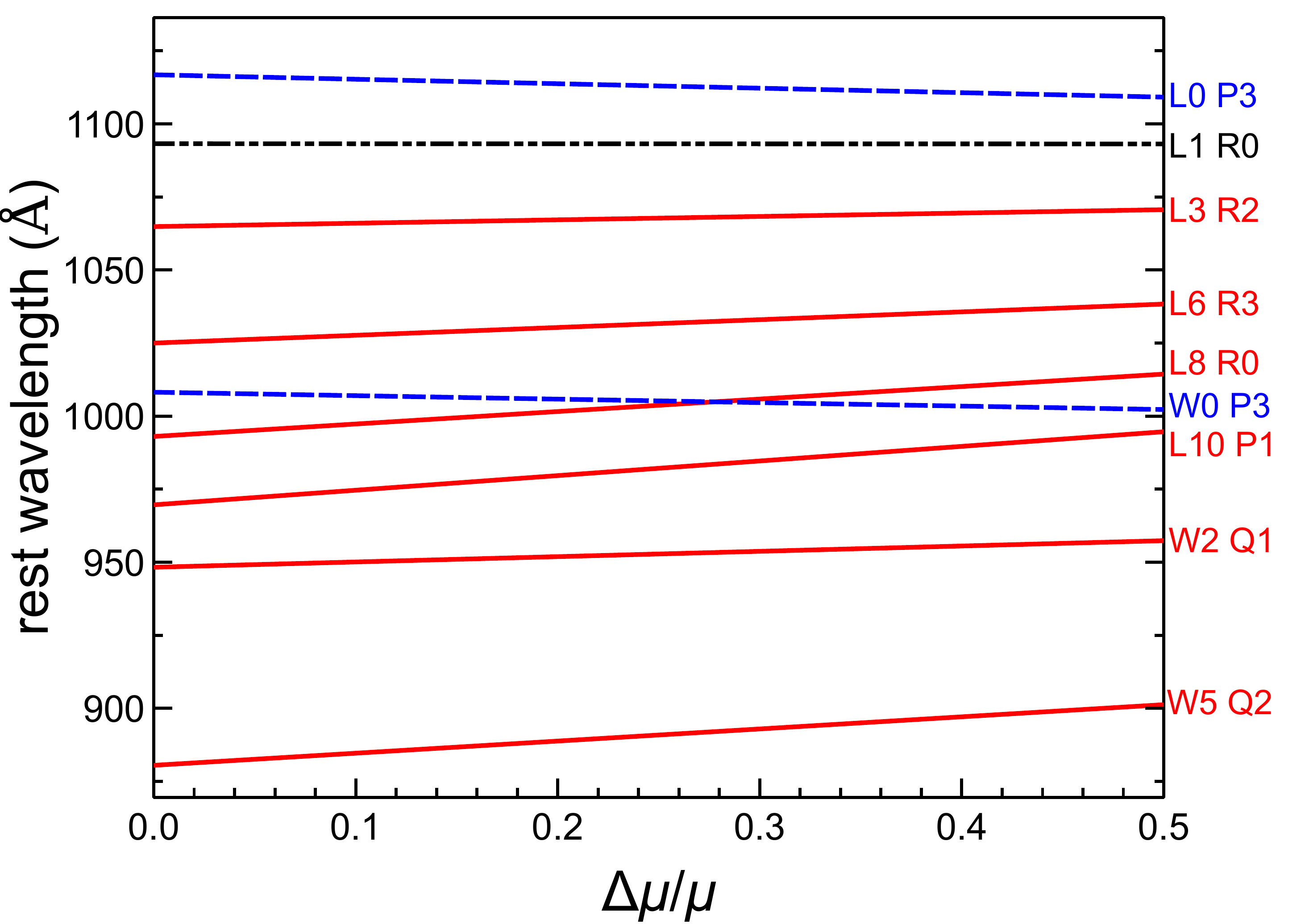}
\caption{Illustration of shifts in the rest wavelengths of Lyman (L) and Werner (W) lines as functions of the (exaggerated) relative variation in $\mu$. The solid (red) curves indicate transitions that are red-shifted with increasing $\Delta\mu/\mu$, while the dashed (blue) curves are blue-shifters. The L1R0 transition, indicated by a dot-dashed (black) curve, represents a line with very low sensitivity to $\mu$-variation, i.e. it acts as an anchor line.}
\label{Shifters}
\end{figure}

As has been pointed out by \citet{Thompson1975} the value of $\mu$ influences the pattern of the rovibronic transitions in molecular hydrogen (H$_2$). A small and relative variation of the proton-to-electron mass ratio
\begin{equation}
\frac{\Delta\mu}{\mu} =\frac{\mu_{z}-\mu_0}{\mu_0},
\end{equation}
will then give rise to differential shifts of the individual absorption lines, as displayed in graphical form and vastly exaggerated in Fig.~\ref{Shifters}.
Here $\mu_{z}$ refers to the proton--electron mass ratio in a distant extra-galactic system at redshift $z$ and $\mu_0$ is a reference laboratory value, i.e. a measurement at zero redshift.
A positive $\Delta\mu/\mu$ indicates a larger $\mu$ in the distant system as compared to the laboratory value. The figure shows that some lines act as anchor lines, not being sensitive to a variation of $\mu$. Most lines in the spectrum of H$_2$ act as redshifters, so producing a longer wavelength for a higher value of $\mu$. Only a minor fraction of lines act as blue shifters, like the lines indicated with L0P3 and W0P3, corresponding to the P(3) line of the $B^1\Sigma_u^+ - X^1\Sigma_g^+$ (0,0) Lyman band, and the P(3) line of the $C^1\Pi_u - X^1\Sigma_g^+$ (0,0) Werner band.
Here a four-digit shorthand writing is used for the H$_2$ transitions observed in cold clouds in the line-of-sight toward quasars, like `LXPY' and `WXRY', where L and W refer to the Lyman and Werner bands, `X' denotes the vibrational quantum number of the excited state, P and R (or Q) refer to the rotational transition and `Y' indicates the ground state rotational level probed.

Besides the very small differential shifts caused by a possible variation of $\mu$, the H$_2$ spectral lines observed from distant galaxies undergo a strong redshift due to the cosmological expansion of the universe. Crucially, General Relativity predicts that that all wavelengths undergo a similar redshift, hence
$\lambda_z=(1+z)\lambda_0$ throughout the spectrum. This assumption that redshift is dispersionless underlies all analyses of varying constants in the early universe based on spectroscopic methods.

When combining the effects of cosmological redshift and a small additional effect due to a variation of the proton--electron mass ratio, the wavelength of the $i^{th}$ transition observed in an absorbing system at redshift $z_{\rm{abs}}$ is then defined as:
\begin{equation}
\lambda^z_i= \lambda^{\rm{0}}_i (1+z_{\rm{abs}})(1+K_i\frac{\Delta\mu}{\mu}),
\label{comp-equation}
\end{equation}
where $\lambda_{\rm{0}}$ is the corresponding rest wavelength, and $K_i$ is a sensitivity coefficient, which will be discussed in section~\ref{section-K}.

In the comparison between H$_2$ absorption spectra recorded at high redshift and laboratory spectra, as much as possible known molecular physics is implemented when deriving a putative effect on $\Delta\mu/\mu$. The H$_2$ molecular lines are determined by a number of parameters, for each transition: (i) the rest-frame wavelength $\lambda_i^0$ which is determined in high-precision laser-based laboratory studies (see section~\ref{section-laser});  (ii) the sensitivity coefficient $K_i$ (see section~\ref{section-K}); (iii) the intensity, rotational linestrength or oscillator strength $f_i$; (iv) the damping factor or natural line broadening factor $\Gamma_i$ (see section~\ref{H2-database}). In the analysis of astronomical spectra this molecular physics information is combined with environmental information to determine the effective line positions, line strengths and line widths: (v) the redshift parameter $z$ and, (vi) the Doppler broadening coefficient $b$, and (vii) the column densities $N_J$, which are identical for each subclass of rotational states $J$ populated in the extra-galactic interstellar medium. These seven parameters determine a fingerprint spectrum of the H$_2$ molecule. While the Doppler broadening coefficient $b$ may in many cases be assumed to be the same for all rotational levels, as is expected to result from homogeneous temperature distributions, in some cases $J$-dependent values for $b$ were found~\cite{Noterdaeme2007b,Rahmani2013}. \citet{Malec2010} discussed that this phenomenon can also be ascribed to underlying and unresolved velocity structure on the H$_2$ absorptions.

The partition function of H$_2$ is such that, at room-temperature or lower temperatures as in most extragalactic sources observed (50--100 K), only the lowest vibrational quantum state in the electronic ground state is populated: $X^1\Sigma_g^+, v=0$. Under these conditions the lowest six rotational states $J=0-5$ are probed in the absorption spectra. Vibrational excitation does not play a role at these temperatures.
For a population of only the lower quantum states the onset of H$_2$ absorption is at rest-frame wavelengths of $\lambda_0=1140$~\AA, where the $B-X(0,0)$ band is probed. The absorption spectrum of H$_2$ extends in principle to the ionization limit and beyond~\cite{Chupka1969}, so to wavelengths as short as 700~\AA, but is truncated at 912~\AA~because of the Lyman-cutoff, i.e. the onset of the absorption continuum due to H\,\textsc{i}.

The other constraint on the coverage of the H$_2$ absorption spectrum is related to the atmospheric window and the reflectivity of mirror coatings generally used at large telescopes. The former cutoffs fall at $\lambda_c=3050$ \AA, which means that an H$_2$ spectrum is fully covered until the Lyman cutoff if the redshift of the absorbing galaxy is $z>2.3$. For absorbing systems at redshifts $z<2.3$ the shortest wavelength lines lie beyond the onset of atmospheric opacity and will not contribute to the observable spectrum. All this means that in the best case the Lyman bands of H$_2$ can be followed up to $B-X (18,0)$ and the Werner bands up to $C-X (4,0)$.

\begin{figure}
\centering
\includegraphics[width=1.\columnwidth]{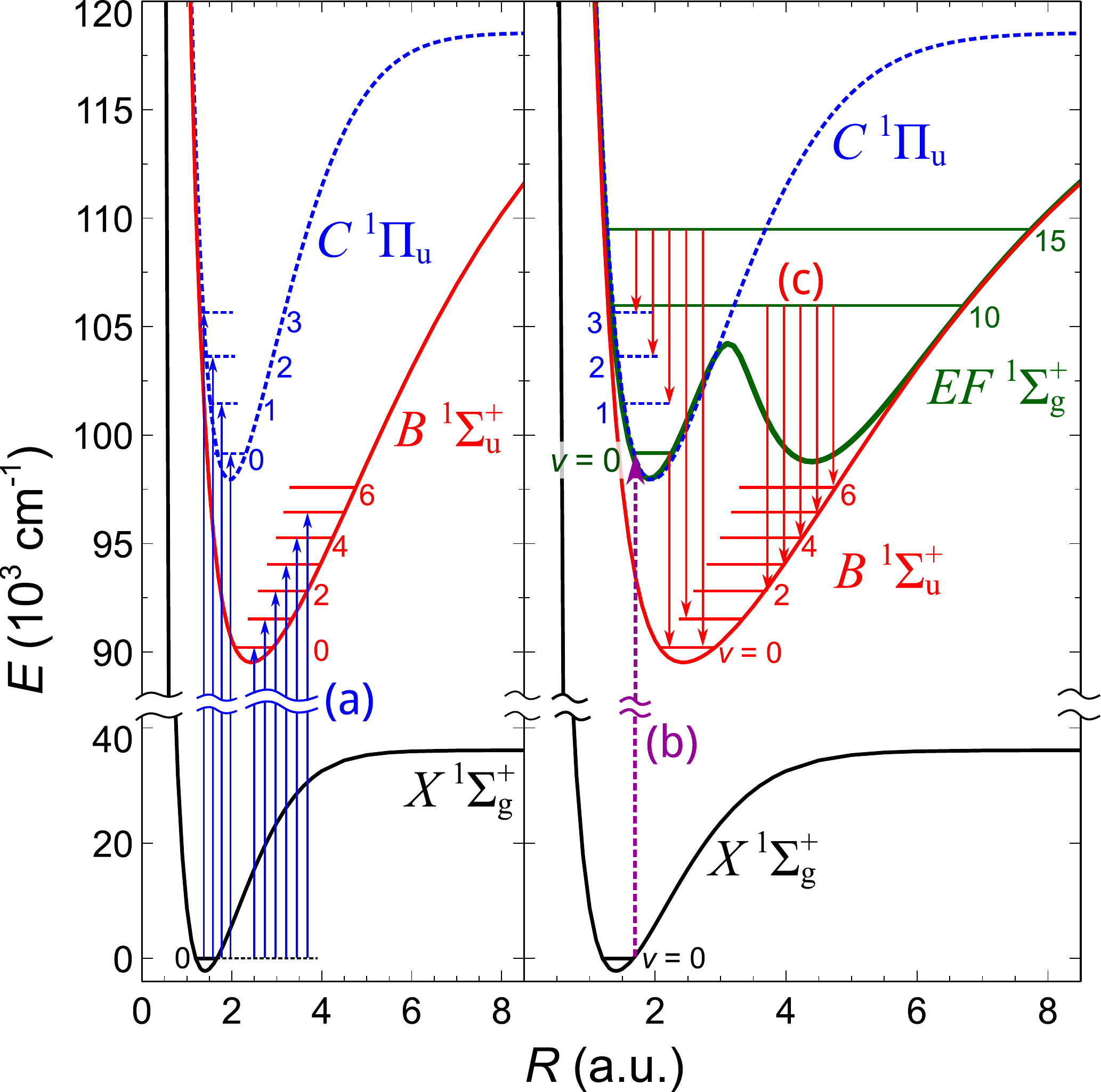}
\caption{
Excitation schemes used to determine the Lyman and Werner transition wavelengths. Represented in the left panel, is a direct excitation (a) scheme from $X$ to $B, C$ states using a narrowband XUV laser system (or alternatively, detection of XUV emission from excited $B$ and $C$ states).
In the right panel, an indirect determination scheme is represented comprising two-photon Doppler-free excitation (b) in the $EF-X$ system and Fourier-transform emission measurements (c) of the $EF-B$ and $EF-C$ systems.}
\label{level-scheme}
\end{figure}

\subsection{Laboratory investigations of molecular hydrogen}
\label{section-laser}

The strongest absorption systems of molecular hydrogen correspond to the electronic excitation of one of the $1s$ ground state orbitals to an excited $2p\sigma$ orbital or to the twofold degenerate $2p\pi$ orbital.
In the former case the $B^1\Sigma_u^+$ molecular symmetry state is probed via the Lyman system ($B^1\Sigma_u^+ - X^1\Sigma_g^+$).
In the latter case the $C^1\Pi_u$ molecular symmetry state is excited in the Werner system ($C^1\Pi_u - X^1\Sigma_g^+$).

For each vibrational band, the rotational series or branches are labeled according to the change in rotational angular momentum
$\Delta J= J' - J''$, where $J'$ and $J''$ are the rotational quantum numbers of the upper and lower levels, respectively.
Thus the  P($J''$) branch is for transitions with $\Delta J = -1$, Q($J''$) is for $\Delta J=0$, and R($J''$) for $\Delta J= +1$. For reasons of molecular symmetry in the Lyman bands only P and R lines occur, while all three types occur in the Werner bands.

There is an extensive data set of emission measurements of the Lyman bands~\cite{Abgrall1993b} and the Werner bands~\cite{Abgrall1993c} that were analyzed with a classical 10m spectrograph. Upon averaging over all emission bands this resulted in values for level energies $B^1\Sigma_u^+, v, J$ and $C^1\Pi_u, v, J$ at an absolute accuracy of $0.15$ \wn~\cite{Abgrall1993a}, which corresponds to $\Delta\lambda/\lambda\approx 1.5\times 10^{-6}$ at $\lambda = 1000$ \AA.
From these data a comprehensive list of absorption lines in the $B-X(v',0)$ Lyman and $C-X(v',0)$ Werner bands, all originating from the ground vibrational level, is calculated. This forms a backup list at an accuracy of $0.15$ \wn\ to be used for those lines where no improved laser-based calibrations have become available.

\begin{figure}
\centering
\includegraphics[width=1.\columnwidth]{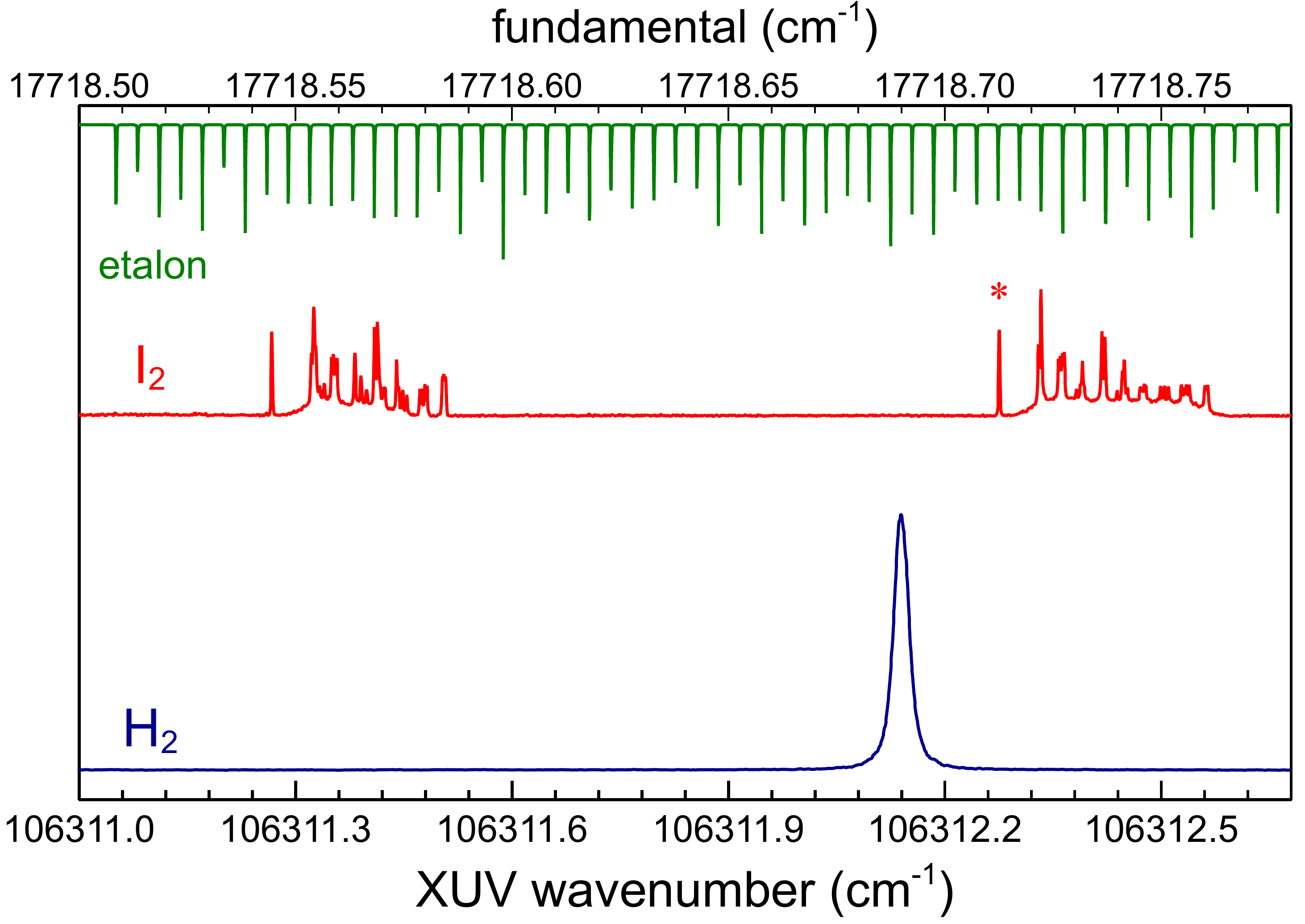}
\caption{Recording of the XUV-absorption spectrum of L15R2, i.e. the R(2) line in the $B^1\Sigma_u^+ - X^1\Sigma_g^+$ (15,0) Lyman band (lower spectrum) with the I$_2$ reference spectrum (middle spectrum)
and \'etalon markers (top spectrum) for the calibration. The line
marked with an asterisk is the a$_1$ hyperfine component of the R(66) line in the $B-X(21,1)$ band in I$_2$ at $17\,718.7123$~\wn.}
\label{XUV-line}
\end{figure}

More recently laser-based laboratory studies of molecular hydrogen were carried out employing direct XUV-laser excitation (see left panel of Fig.~\ref{level-scheme}) of rotational lines in the Lyman and Werner bands of the H$_2$ molecule. The measurements were performed using a narrowband and tunable laser system in the visible wavelength range, which is upconverted via frequency-doubling in crystals and third-harmonic generation in xenon gas jets delivering coherent radiation in the range 900--1150 \AA. The method of 1 + 1 resonance-enhanced multi-photon ionization was employed for detection of the spectral resonances. The spectroscopic resolution was as accurate as 0.02 \wn, achieved via sub-Doppler spectroscopy in a molecular beam. The absolute calibration of the H$_2$ resonances was performed by interpolation of the frequency scale in the visible range by a stabilized etalon and by comparing to a saturated absorption spectrum of molecular iodine~\cite{Velchev1998,Xu2000}. An example of a recording of the L15R2 line is displayed in Fig.~\ref{XUV-line}.

More than $160$ H$_2$ spectral lines in the $B-X(v',0)$ Lyman ($v'$ up to 19) and $C-X(v',0)$ Werner bands ($v'$ up to 4) were calibrated. Only the vibrational ground state $v=0$ was probed and in most cases $J=0-5$ rotational states, although in some cases only the very lowest rotational quantum states. The results, obtained at a typical accuracy of $\Delta\lambda/\lambda = 5 \times 10^{-8}$, have been published in a sequence of papers covering the relevant wavelength range 900-1150 \AA\ ~\cite{Ubachs2004,Philip2004,Reinhold2006,Hollenstein2006,Ivanov2008b}. In cases where only population of the lowest rotational quantum states $J=0-2$ could be probed, mainly for the R-branch lines, the wavelength positions of P($J$) lines were calculated to high accuracy from rotational combination differences in the $X^1\Sigma_g^+, v=0$ ground state, derived from the precisely measured pure rotational spectrum of H$_2$ in the far-infrared~\cite{Jennings1984}.

Similar XUV-laser studies were performed for the HD molecule~\cite{Ivanov2008a}, achieving an accuracy of $\Delta\lambda/\lambda = 5 \times 10^{-8}$, while also a more comprehensive study on HD was performed using VUV-Fourier-transform absorption spectroscopy with synchrotron radiation yielding a lower accuracy of $\Delta\lambda/\lambda = 4 \times 10^{-7}$~\cite{Ivanov2010}.

\begin{figure}
\centering
\includegraphics[width=1.\columnwidth]{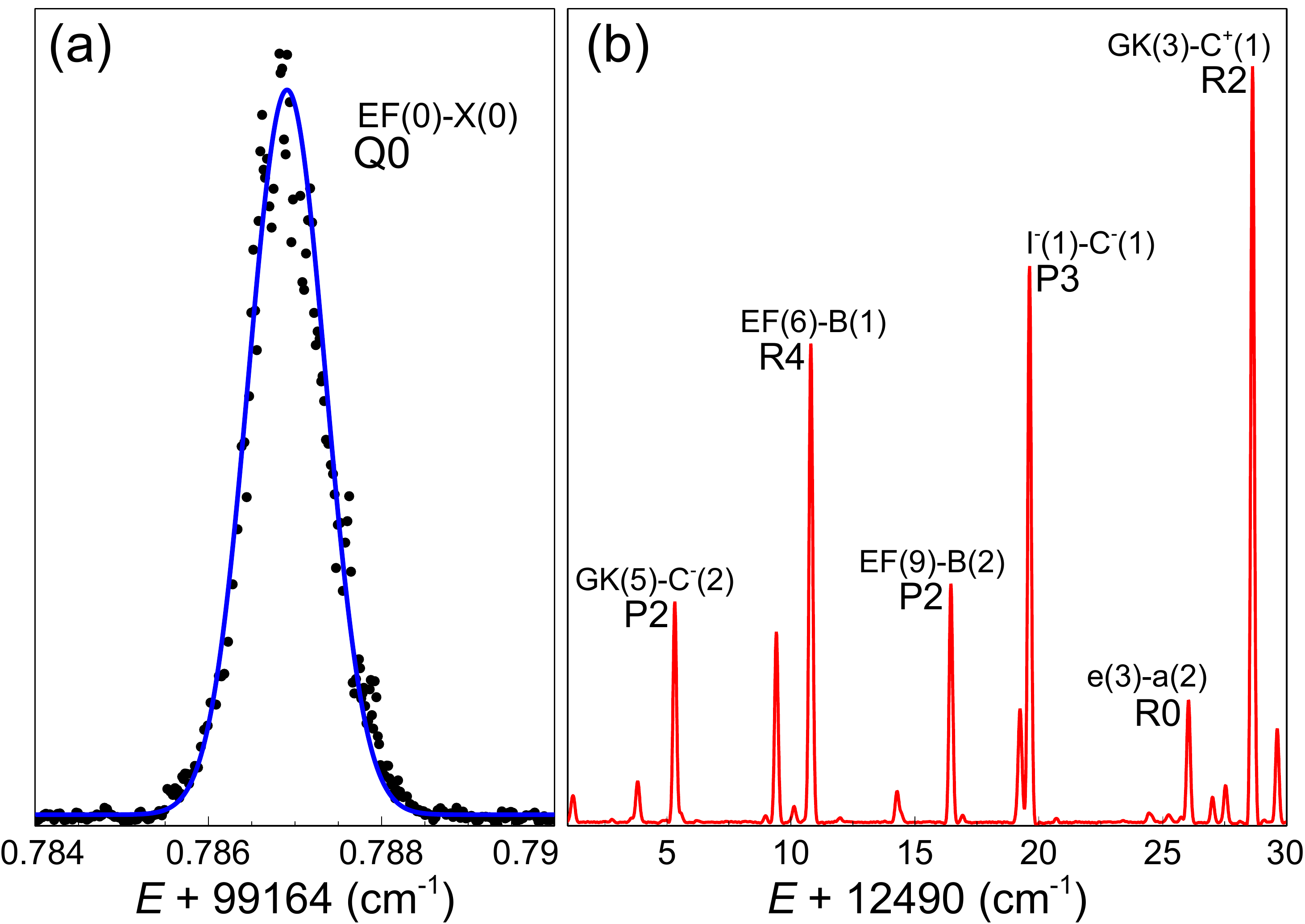}
\caption{(a) Laser spectroscopic recording of the Q(0) two-photon transition in the $EF^1\Sigma_g^+ - X^1\Sigma_g^+$ (0,0)  band; (b)
Recording of part of the Fourier-transform spectrum of the $EF^1\Sigma_g^+ - B^1\Sigma_u^+$ emission. }
\label{VUV-FTIR}
\end{figure}

An improved laboratory wavelength determination of the Lyman and Werner bands is derived from a combination of laser-based deep-UV (DUV) laser spectroscopy and Fourier Transform (FT) spectroscopy  \cite{Salumbides2008,Bailly2010}. The advantage of this two-step method, schematically depicted in the right-hand panel of Fig.~\ref{level-scheme}, is that it bypasses the instrument bandwidth limitation in the direct XUV-laser spectroscopy described above.
In the FT-studies, emission from a low-pressure H$_2$ microwave discharge was sent into an FT spectrometer, and a set of filters and detectors were used to cover the broad wavelength range between 5 $\mu$m and 4500 \AA\ ($2000 - 22\,000$ \wn). A small part of the obtained FT-spectrum is plotted in Fig.~\ref{VUV-FTIR}(b).
The comprehensive FT spectra include transitions belonging to different vibrational bands of (symmetry-allowed) pair combinations of electronic states: $B\,^1\Sigma^+_u$, $C\,^1\Pi_u$, $B'\,^1\Sigma^+_u$, $EF\,^1\Sigma^+_g$, $GK\,^1\Sigma^+_g$, $H\,^1\Sigma^+_g$, $I\,^1\Pi_g$, and $J\,^1\Delta_g$.
The accuracy of the line position determination varies for the different wavelength ranges, ultimately limited by Doppler broadening. For strong lines in the infrared range accuracies of 0.0001 \wn\  could be obtained.

To connect the level structure of the excited electronic state manifold to the $X^1\Sigma^+_g\, v=0, J=0$ level, results from Doppler-free two-photon spectroscopy on the $EF-X$ (0,0) band are used~\cite{Hannemann2006,Salumbides2008}.
In these laser-spectroscopic investigations, a pulsed narrowband laser source, frequency-comb assisted frequency calibration, interferometric alignment for Doppler-shift suppression, frequency-chirp measurement, and ac-Stark correction, are all essential ingredients to achieve the $\sim10^{-9}$  accuracies~\cite{Hannemann2006}.
The Q(0) and Q(1) lines, with accuracies better than $2\times10^{-4}$ \wn, are specifically used as anchor lines for para- and ortho-H$_2$, respectively.
A recording of the Q(0) line is plotted in Fig.~\ref{VUV-FTIR}(a).
A comparison of the Q($J''=2-5$) transitions measured in the DUV investigations with those derived from the FT investigations confirmed the accuracy of the two-step method of obtaining the Lyman ($\Delta\lambda/\lambda \sim 10^{-9}$) and Werner ($\Delta\lambda/\lambda \sim 10^{-8}$) transition wavelengths.

\subsection{Sensitivity coefficients}
\label{section-K}

In comparing astrophysical spectra, yielding the wavelengths at high redshift $\lambda_i^z$, with laboratory spectra providing $\lambda_i^0$ a fit can be made to extract or constrain a value of $\Delta\mu/\mu$ via Eq.~(\ref{comp-equation}).
Here $K_i$ is the sensitivity coefficient, different for each transition, and defined by:
\begin{equation}
  K_i = \frac{d\ln \lambda_i}{d \ln \mu}
\label{sens-coeff}
\end{equation}
This equation can also be expressed in terms of the quantum level energies $E_g$ and $E_e$ of ground and excited states involved in a transition and a spectroscopic line:
\begin{equation}
  K_i =  - \frac{\mu}{E_e - E_g} \left[ \frac{dE_e}{d\mu} - \frac{dE_g}{d\mu}  \right].
\label{sens-E}
\end{equation}
It is noted that some authors calculate sensitivity coefficients to a possible variation of the electron--proton mass ratio $\mu'=m_e/m_p=1/\mu$, in which case the values of the $K_i$ reverse sign \cite{Kozlov2013b}. Also sometimes the sensitivity coefficients are defined in connection to frequencies \cite{Jansen2014}, $\Delta\nu/\nu = K\Delta\mu/\mu$, in which case the values of the $K_i$ also reverse sign.

The sensitivity coefficient $K_i$ can be considered as an isotope shift of a transition in differential form. Its value can in good approximation be understood in the Born-Oppenheimer approximation, separating the contributions to the energy of the molecule
\begin{equation}
E_{i} = E_{\rm{elec}} + E_{\rm{vibr}}(\mu^{-1/2}) + E_{\rm{rot}}(\mu^{-1}),
\end{equation}
with $E_{\rm{elec}}$ the electronic, $E_{\rm{vibr}}$ the vibrational, and $E_{\rm{rot}}$ the rotational energy. The dependence of these energy terms on $\mu$ is then known: (i) in the Born--Oppenheimer approximation the electronic energy is independent of $\mu$; (ii) in a harmonic oscillator approximation the vibrational energy scales as $(1/\sqrt{\mu})$; (iii) in a rigid rotor approximation the rotational energy scales as $1/\mu$.
These scalings explain the values of the sensitivity coefficients. For the (0,0) bands in the Lyman $B-X$ and Werner $C-X$ systems the transition is almost fully electronic in nature and henceforth $K_i \approx 0$. For $(v',0)$ bands probing higher vibrational energies a fraction of the excited state energy becomes vibrational in nature and therefore the value of $K_i$ increases. Because the amount of vibrational energy remains below 15\% the value of $K_i$ will remain below 0.07, where it is noted that for pure vibrational energy $K_i = -1/2$. Similarly for the fraction of rotational energy $K_i = -1$ \cite{Jansen2014}.

The $K$-values can be derived in a semi-empirical approach, separating electronic energies and expressing rovibrational energies in a Dunham representation:
\begin{equation}
E(v,J) = \sum_{k,l} Y_{k,l} \left(v + \frac{1}{2} \right)^k  [J(J+1) - \Lambda^2]^l
\end{equation}
with Dunham coefficients $Y_{k,l}$, $v$ and $J$ vibrational and rotational quantum numbers, and $\Lambda$ signifying the electronic orbital momentum of the electronic state (0 for the $X^1\Sigma_g^+$ and $B^1\Sigma_u^+$ states and 1 for the $C^1\Pi_u$ state). This method draws from the advantage that the energy derivatives $dE/d\mu$ in Eq.~(\ref{sens-E}) can be replaced by derivatives $dY/d\mu$ and that the functional dependence of the Dunham coefficients $Y_{k,l}$ is known~\cite{Reinhold2006,Ubachs2007}.
\citet{Varshalovich1993} were the first to calculate $K$-coefficients based on such a Dunham representation of level energies of ground and excited states. Thereafter \citet{Ubachs2007} recalculated $K$-coefficients from improved accuracy laser-based measurements of Lyman and Werner bands within the Dunham framework.  They also showed how to correct the $K$-values for adiabatic and non-adiabatic effects in the excited states.

Alternatively, numerical values for the sensitivity coefficients can be derived by making use of \emph{ab initio} calculations for the hydrogen molecule, as was pursued by~\citet{Meshkov2006}, yielding values $K_{AI}$. These are found to be in good agreement with the values $K_{SE}$ obtained in the semi-empirical approach~\cite{Ubachs2007}, within an accuracy of $\Delta K = |K_{AI} - K_{SE}| < 3 \times 10^{-4}$.

\begin{figure}
\centering
\includegraphics[width=1.\columnwidth]{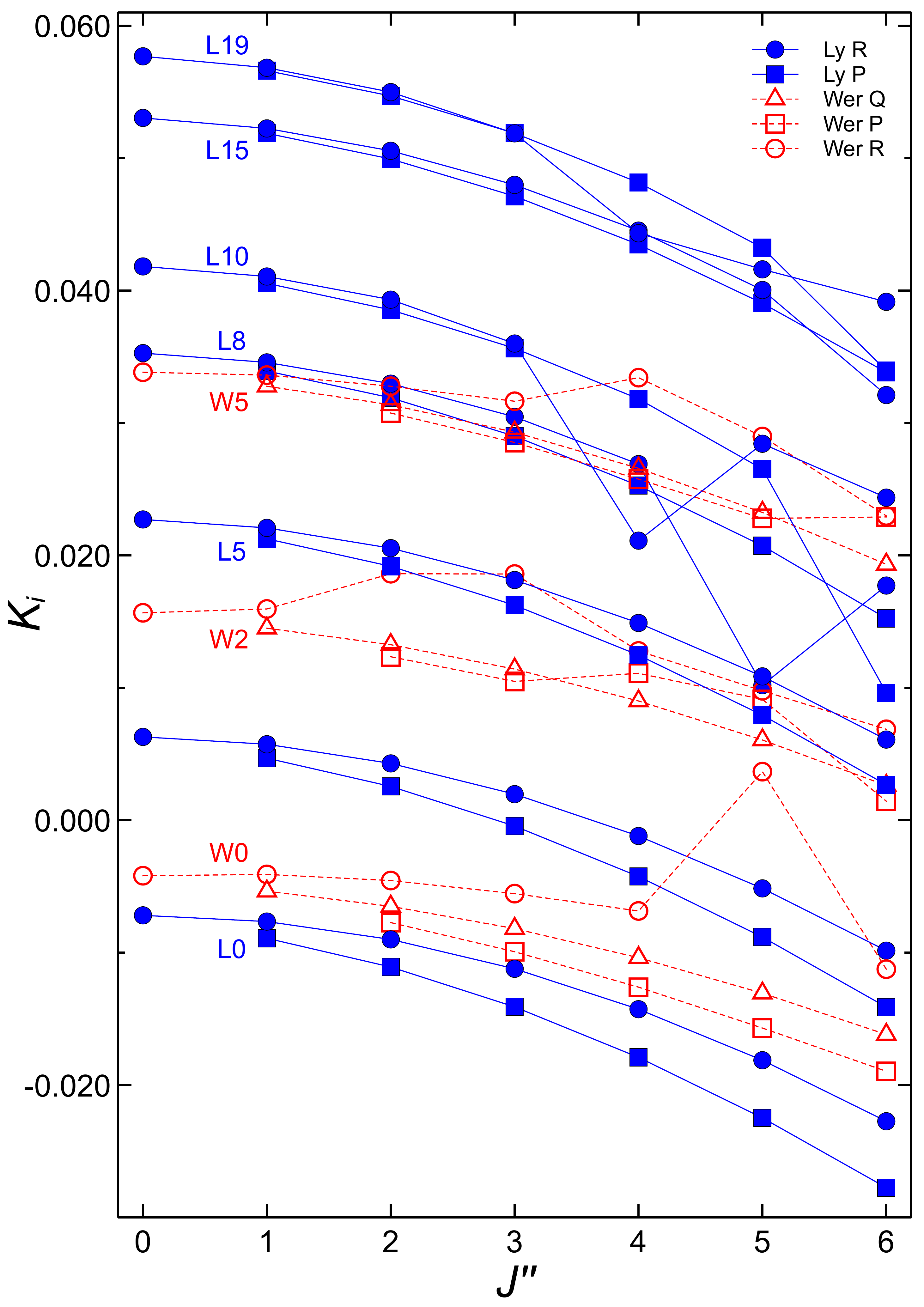}
\caption{
Sensitivity coefficients $K_i$ of individual lines in the Lyman and Werner bands, showing the range of values from around -0.02 to 0.06.
Filled (blue) symbols are associated with Lyman lines and open (red) symbols to Werner lines.
From L8, level crossings occur between levels in the $B\,^1\Sigma^+$ and $C\,^1\Pi^{+}$ states, leading to irregularities in the $K$ progression of the Lyman P, R and Werner Q transitions.
}
\label{K-coeffs}
\end{figure}

Recently, improvements were made following both the semi-empirical and the \emph{ab initio} approaches~\cite{Bagdonaite2014b}.
For the semi-empirical analysis it was realized that fitting the Dunham coefficients is not necessary. Instead, derivatives of the level energies $dE/d\mu$ can be obtained from numerical partial differentiation with respect to the vibrational $v$ and rotational $J$ quantum numbers
\begin{equation}
	\frac{dE}{d\mu}\Bigg|_{v,J} = -\frac{1}{2\mu} \left( v + \frac{1}{2} \right) \frac{\partial E}{\partial v}\Bigg|_{v,J} - \frac{1}{\mu} \frac{J(J+1)}{2J+1} \frac{\partial E}{\partial J}\Bigg|_{v,J}.
\label{SE-Edcel}
\end{equation}
This provides a more direct procedure only requiring derivatives to calculate sensitivity coefficients $K_i$, while in practice the calculation of derivatives appears to be more robust than calculating strongly correlated Dunham coefficients.

Further, an improved round of \emph{ab initio} calculations were carried out including the best updated numerical representations of the four interacting excited state potentials for $B^1\Sigma_u^+$,  $B'^1\Sigma_u^+$, $C^1\Pi_u$, and $D^1\Pi_u$ states \cite{Staszewska2002,Wolniewicz2003}, including adiabatic corrections and the mutual non-adiabatic interactions~\cite{Wolniewicz2006}. These calculations were performed for a center value for the reduced mass of $\mu_{red}= 0.5\mu$ for H$_2$ and ten different intermediate values following an incremental step size for $\Delta\mu/\mu$ of 10$^{-4}$ around the center value. Taking a derivative along the $\mu$-scale yields \emph{level sensitivity} coefficients $\mu dE/d\mu$ for the excited states. A similar procedure was followed for the $X^1\Sigma_g^+$ ground state based on the Born-Oppenheimer potential computed by \citet{Pachucki2010} and non-adiabatic contributions of \citet{Komasa2011}. Accurate values for sensitivity coefficients were derived by dividing through the transition energies, as in the denominator of Eq.~(\ref{sens-E}).

Values for the $K_i$ sensitivity coefficients pertaining to transitions originating in the lowest rotational quantum states are plotted in Fig.~\ref{K-coeffs} for a selected number of Lyman and Werner bands. These values obtained from recent \emph{ab initio} calculations are in very good agreement with the updated semi-empirical calculation employing Eq.~(\ref{SE-Edcel}). The negative values for W0 and L0 bands, due to the fact that the zero-point vibration in the ground state is larger than in the excited states, makes those lines blue-shifters: they shift to shorter wavelengths for increasing $\mu$. The irregularities in the data progressions, such as for the R(5) line in the W0 band, show the local mutual (non-adiabatic) interactions between $B^1\Sigma_u^+$ and $C^1\Pi_u$ states, giving rise to state mixing. The lines in the L19 band exhibit the largest $K$-coefficients, because the fraction of vibrational energy to the total energy in the excited state is largest. Higher vibrational bands than L19 in the Lyman system are not considered because they are at a rest wavelength $\lambda < 910$ \AA\ and therewith beyond the Lyman cutoff of atomic hydrogen. Sensitivity coefficients for the HD molecule were also calculated through the \emph{ab initio} approach~\cite{Ivanov2010}.

\begin{figure}[!ht]
\centering
\includegraphics[width=0.95\columnwidth]{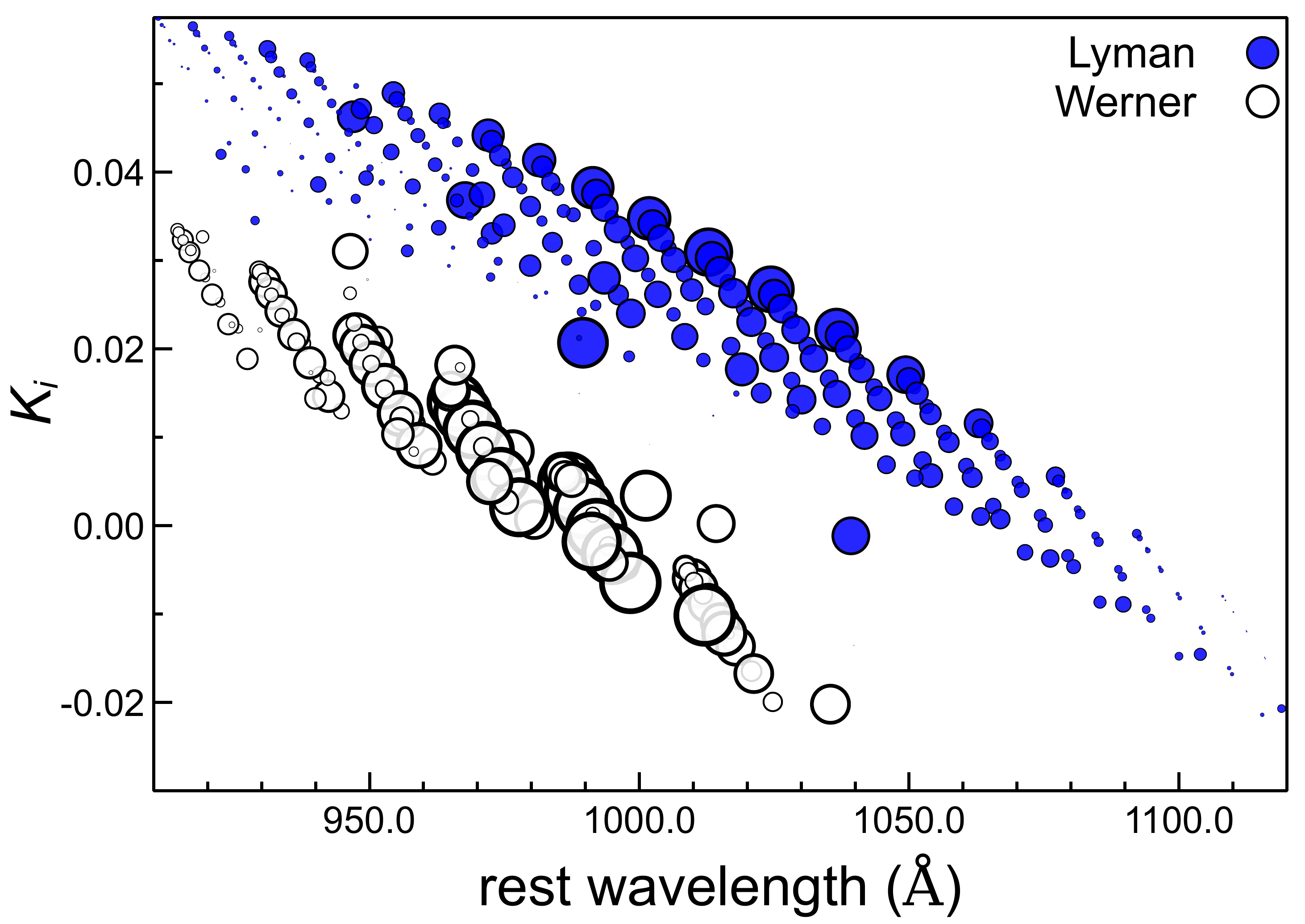}
\caption{Sensitivity coefficients $K_i$ for $\mu$-variation in H$_2$ as a function of rest-frame wavelengths.
Filled (blue) symbols are associated with Lyman lines and open symbols to Werner lines.
The size of the data points reflects line oscillator strengths $f_i$ for the spectral lines.}
\label{KMu-strength}
\end{figure}

In Fig.~\ref{KMu-strength} the sensitivity coefficients to $\mu$-variation are plotted as a function of wavelength.
The size of the data points reflects the line oscillator strength $f_i$ for the H$_2$ lines, hence representing their significance in a $\mu$-variation analysis.
The significance of the various spectral lines in a $\mu$-variation analysis depends further on the population distribution in the absorbing cloud as well as on the total column density. A typical population temperature is $ \approx 50-100$ K, and therefore the population of H$_2$ is dominantly in the $J=0$ and $J=1$ para and ortho ground states.
However, the population of rotational quantum states of H$_2$ in the absorbing gas is likely to depart from thermodynamic equilibrium, so that the population distribution reflects some combination of the prevailing kinetic temperature, the radiation environment in the absorbing region, and the rates of chemical processes involving H$_2$.
This is contrary to the case of CO which is found to reflect a partition function commensurate with the local cosmic background temperature $T_{\rm{CO}}= (1+z) T_{\rm{CMB}}$ \cite{Noterdaeme2011}. HD is so far detected only in the $J=0$ state~\cite{Tumlinson2010,Ivanchik2010,Noterdaeme2010}.
In case of very large column densities, i.e.~for log$N$[H$_2$/cm$^{-2}] > 18$, the low-$J$ levels will be saturated and broadened, making them less suitable to determine line centers. At these high column densities the spectral lines with $J \geq 1$, and even with $J \geq 3$ will most decisively contribute to constraining $\Delta\mu/\mu$.

The distribution of points in Fig.~\ref{KMu-strength} demonstrates that $K$-coefficients correlate strongly with wavelength for each band system separately. Observation of both Werner and Lyman systems counteracts the wavelength correlation of the $K$-values and makes the analysis more robust.

\subsection{The H$_2$ molecular physics database}
\label{H2-database}

The extraction of information from high-redshift observations of H$_2$ molecular spectra depends on the availability of a database including the most accurate molecular physics input for the relevant transitions in the Lyman and Werner absorption bands. Values for the wavelengths $\lambda_i$ were collected from the classical data~\cite{Abgrall1993b,Abgrall1993c}, the direct XUV-laser excitation~\cite{Reinhold2006,Philip2004,Hollenstein2006,Ivanov2008b}, and the two-step excitation process~\cite{Salumbides2008,Bailly2010}; in each case the most accurate wavelength entry is adopted. The $K_i$ sensitivity coefficients are included from the recent \emph{ab initio} calculations involving a four-state interaction matrix for the excited states~\cite{Bagdonaite2014b}. In order to simulate the spectrum of molecular hydrogen also values for the line oscillator strength $f_i$ and radiative damping parameter $\Gamma_i$, giving rise to Lorentzian wings to the line shape, should be included for all H$_2$ lines in the Lyman and Werner bands. These were calculated by~\citet{Abgrall2000}. The recommended data base containing the molecular physics parameters for H$_2$ and HD needed to model quasar absorption spectra was published as a supplementary file by~\citet{Malec2010}.

\section{Astronomical observations of H$_2$}
\label{astronomy}

\begin{figure*}
\centering
\includegraphics[width=1.\textwidth]{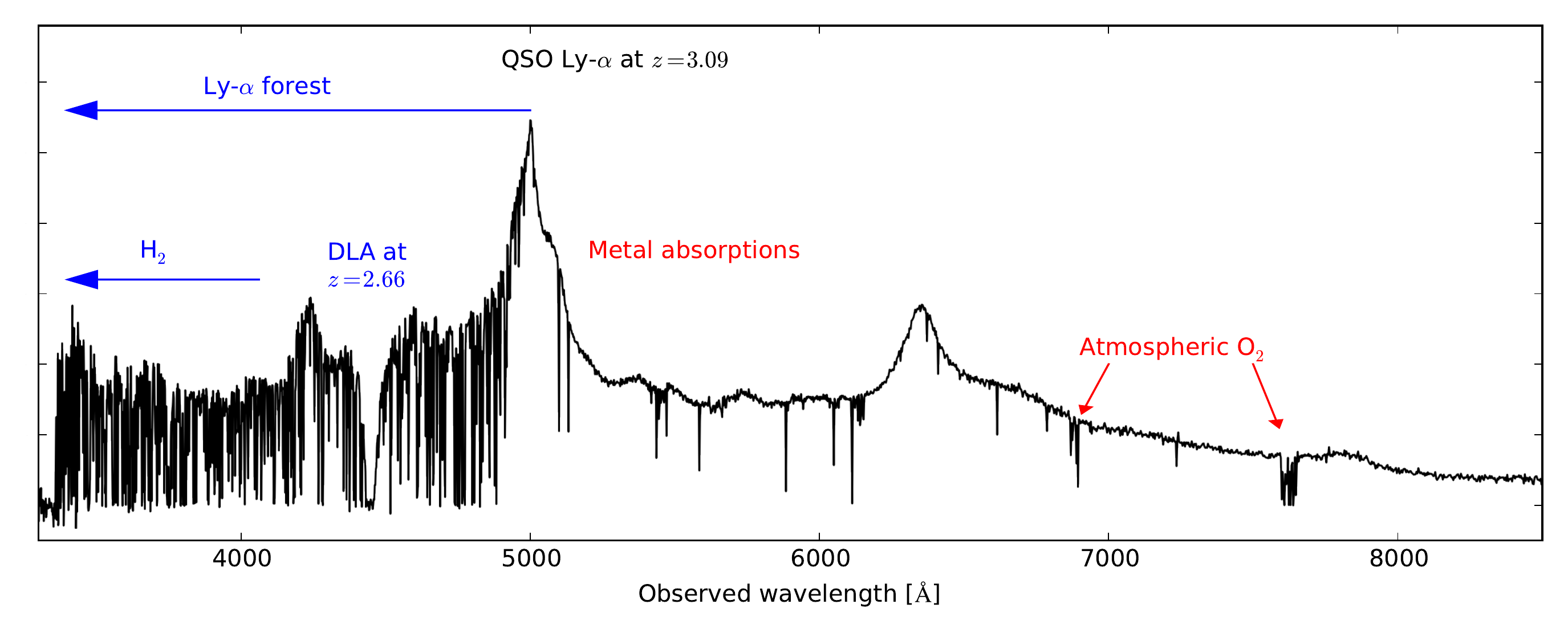}
\caption{Typical spectrum of a quasar displaying the observed flux (in arbitrary units), in this case for B0642$-$5038 emitting at redshift $z=3.09$ and registered by the Ultraviolet and Visual Echelle Spectrograph (UVES) mounted at the ESO Very Large Telescope (VLT). The quasar spectrum displays the characteristic broad emission line profiles produced by e.g.~Lyman-$\alpha$ of H\,$\textsc{i}$, and C\,\textsc{iv}. The Lyman-$\alpha$ forest arises as the light from the quasar crosses multiple neutral hydrogen clouds lying in the line of sight. At the blue side of the emission peak the damped Lyman-$\alpha$ system at $z_{\rm{DLA}}=2.66$ causes a series of very strong absorption lines starting at $\lambda=(z_{\rm{DLA}}+1)\times$1216 \AA\,= 4450~\AA, and absorbs all the flux at $\lambda <(z_{\rm{DLA}}+1)\times$912 \AA\,= 3340~\AA~producing a so-called Lyman break or Lyman limit cutoff. Since the rest wavelengths of H$_2$ are in the UV ($<1140$ \AA) they are found in the Lyman-$\alpha$ forest. At the red side of the Lyman-$\alpha$ emission peak the sharp absorption lines are due to metallic ions (C\,\textsc{iv}, Fe\,\textsc{ii}, Si\,\textsc{iv}, etc.) at various redshifts including that of the DLA. The absorptions at 6800 \AA\ and at 7600 \AA\ are related to the Fraunhofer B and A bands (molecular oxygen) absorbing in the Earth's atmosphere.}
\label{fig-spectrum}
\end{figure*}

While the earliest observations of H$_2$ at high redshift along quasar sight-lines were performed in the 1970s~\cite{Carlson1974,Aaronson1974} the first study of $\mu$-variation using the H$_2$ method were carried out somewhat later by~\citet{Foltz1988} using the Multiple Mirror Telescope (Arizona) composed of 6 dishes of 1.8m.
When the 8--10m class telescopes became available, such as the Keck telescope in Hawaii and the ESO Very Large Telescope (VLT) at Paranal (Chile) both equipped with high resolution spectrographs,  systematic investigations of H$_2$ absorption systems at intermediate and high redshifts have been pursued.

An example quasar spectrum of the source B0642$-$5038, recorded with the Ultraviolet/Visible Echelle Spectrograph (UVES) mounted on the VLT~\cite{Noterdaeme2008} (raw data available in the ESO archive) is displayed in Fig.~\ref{fig-spectrum}. A characteristic feature of such a spectrum is the strong and broad central H\,$\textsc{i}$ Lyman-$\alpha$ emission peak at 4980\,\AA, yielding a value for the redshift of the quasar ($z=3.09$). Also a weaker emission peak related to Lyman-$\beta$ is found near 4250\,\AA, as well as a C\,\textsc{iv} emission peak at 6300\,\AA. The broad absorption feature exhibiting Lorentzian wings at 4450\,\AA\ is the `damped Lyman-$\alpha$ system' (DLA) found at redshift $z=2.66$. DLAs are thought to mainly arise in distant, gas-rich galaxies, on the verge of producing bursts of star formation~\cite{Wolfe2005}. The redshift of the DLA also determines the wavelength of the 13.6\,eV `Lyman-limit' cutoff in the spectrum, in this case at 3340\,\AA. Shortward of this wavelength all radiation is absorbed in the DLA. The `Lyman-$\alpha$ forest' -- the collection of narrower absorption lines covering the entire region between the Lyman-cutoff and the Lyman-$\alpha$ emission peak -- arises mainly in the lower-density, filamentary structures of the intergalactic medium traced by diffuse neutral hydrogen~\cite{Meiksin2009}. The forest absorption lines are (close to) randomly distributed in redshift, giving their forest-like appearance in the spectrum. The region bluewards of the Lyman-$\beta$ emission peak also includes the Lyman-$\beta$ absorption lines corresponding to the forest lines at higher redshift, and so on for successive emission peaks of the hydrogen Lyman series. Of importance for the search of a varying $\mu$ is the location of the H$_2$ absorption lines. These molecular transitions are related to the denser DLA cloud at redshift $z=2.66$, and hence the wavelength region of Lyman and Werner H$_2$ absorption is $(1+z_{\rm{DLA}})\times [910$--1140]\,\AA, or [3340--4170]\,\AA\ in the example shown in Fig.~\ref{fig-spectrum}.

\subsection{Spectral observations and instrumental effects}

The visible (i.e.~redshifted UV) observations of H$_2$ towards background quasars that have so far been used to constrain variations
in $\mu$, have all been obtained using slit-fed, `echelle' spectrographs, the Ultraviolet and Visible Echelle Spectrometer (UVES)
at the VLT \cite{Dekker2000} and the HIRES spectrometer at Keck \cite{Vogt1994}. After passing through a
simple slit, the quasar light is collimated onto the main dispersive element, the `echelle' grating, at grazing incidence.
Most of the diffracted light is in high diffraction orders ($\sim$100), each with a small spectral range ($\sim$60\,\AA)
but high resolving power of typically $R\equiv\lambda/{\rm FWHM} \sim 50,000$ (for FWHM the
full-width-at-half maximum resolution).
To record a wide total spectral range in a single exposure, the diffracted echelle orders are
`cross-dispersed', perpendicular to the echelle dispersion direction, so that many orders can be simultaneously imaged onto a rectangular
charge-coupled device (CCD) whose width exceeds the spatial extent of their spectral range.

Observations at the Keck telescope are normally undertaken by the investigators of the proposed observations themselves. This `visitor mode' is sometimes also used at the VLT. However, most quasar observations at the VLT are taken in `service mode': a local astronomer conducts the observations and calibrations, and the data are delivered to an archive where the original investigators or, increasingly, other researchers can reduce it from raw to usable form with software supplied by the observatory or used-made improved versions of it~\cite{Thompson2009a,Murphy2007}. Many of the quasar observations used for $\mu$-variation studies so far were taken at the VLT in service mode for other purposes.

Depending on the quasar brightness, observations typically involve $\sim$10 separate exposures, of 1--2\,hr duration each, spread over several observing nights under varying temperature, pressure, sky conditions and telescope positions. A variety of calibration exposures are therefore taken at regular intervals to allow these effects and the spectrograph's configuration to be removed from the recorded spectra. For varying-$\mu$ studies, the most important aspect to calibrate accurately is the wavelength scale of each quasar exposure.

The standard wavelength calibration is done via comparison with an exposure of a thorium--argon (ThAr) hollow-cathode emission-line lamp taken in the same spectrograph configuration. The ThAr lamp is mounted within the spectrograph and its light illuminates the slit via a fold mirror moved into position to mimic the light path of the quasar photons through the slit. The ThAr line wavelengths are known from laboratory studies~\cite{Lovis2007}, so a low-order polynomial relationship between wavelength and pixel position on the CCD can be established by fitting the observed lines \cite{Murphy2007}. This relationship is simply assumed to apply to the corresponding quasar exposure. However, two main differences in the light path are cause for concern \cite{Murphy2001,Murphy2003}: the ThAr light may not be aligned with the quasar light closely enough, and the ThAr light fully illuminates the slit whereas the quasar is a point source whose image is blurred to a Gaussian-like spatial distribution by atmospheric `seeing' (refraction and turbulence effects). These effects, particularly the former, may be expected to distort the ThAr wavelength scale away from the true scale for exposures of astronomical objects over long wavelength ranges, so they are important to identify and correct if possible.

Given that the Keck and VLT spectrographs are not actively stabilised, it is therefore important that the ThAr exposure immediately follows the quasar exposure, with no intervening changes to the spectrograph's configurations (especially grating positions). This is referred to as an ``attached'' ThAr. This is generally true for visitor-mode observations but, in service mode at the VLT, this is not the default operation and grating positions are normally re-initialized between the quasar and ThAr calibration exposure; see discussion in \citet{Molaro2013}.

By comparing VLT observations of reflected sunlight from asteroids with solar line atlases, \citet{Molaro2008}
found that the ThAr wavelength scale was accurate enough for varying-constant studies. A similar calibration technique using quasars or stars observed through an iodine gas cell identified substantial distortions of the wavelength scale on scales associated with echelle orders \cite{Griest2010,Whitmore2010}. However, these `intra-order' distortions were repeated from order to order, without any evidence of a long-range component. Therefore, H$_2$ transitions would have effectively random shifts added between them, but no significant systematic effect should be seen in analyses using large numbers of transitions, like the varying $\mu$ work~\cite{Murphy2009,Whitmore2010}.

Unfortunately, more recent investigations using archival VLT and Keck spectra of asteroids and stars with solar-like spectra -- `solar twins' -- found that substantial long-range distortions in ThAr wavelength scales are actually the norm \cite{Rahmani2013,Whitmore2015}; the earlier works appear simply to have sampled unrepresentative periods when these distortions were very small. The effect of these long-range distortions on $\mu$-variation measurements can be understood from Fig.~\ref{KMu-strength}: the $K$ coefficients of the Lyman-band transitions decrease systematically towards longer wavelengths. Therefore, long-range distortions of the quasar wavelength scale will cause a shift in $\Delta\mu/\mu$. The recent investigations of $\mu$-variation with VLT and Keck spectra have shown these long-range distortions to be the main systematic uncertainty in measuring $\Delta\mu/\mu$ accurately. Several of these analyses have been able to use this `supercalibration' information to correct the distortions in the spectra, with varying degrees of confidence due to lack of knowledge about how the distortions change on short time-scales (or spectrograph and telescope configurations), e.g.~the studies of HE0027$-$1836 \cite{Rahmani2013}, B0642$-$5038 \cite{Bagdonaite2014a}, J1237$+$064 \cite{Dapra2015}, and J1443$+$2724 \cite{Bagdonaite2015}. Virtually all the monitored long-range wavelength distortions for the VLT caused a positive shift in the value of $\Delta\mu/\mu$ by a few 10$^{-6}$ (i.e.~correcting for the distortions gave a more negative $\Delta\mu/\mu$).

\subsection{Fitting of spectra}

A first method, referred to as a line-by-line or reduced redshift method, isolates narrow regions in the spectrum where single H$_2$ absorption lines occur, without any or very little visible overlap from Lyman-$\alpha$ forest and metal line absorptions~\cite{Ivanchik2005,Reinhold2006,Thompson2009,Wendt2011,Wendt2012}. Peak positions are then fitted to determine an observed redshift \mbox{$z_i=(\lambda_i^z/\lambda_i^0)-1$} for each transition which is then implemented in a fitting routine, minimizing deviations
to Eq.~(\ref{comp-equation}) and determining values for $\Delta\mu/\mu$ and $z_{\rm{abs}}$.
After the fit the reduced redshift,
\begin{equation}
  \zeta_i = \frac{z_i - z_{\rm{abs}}}{1 + z_{\rm{abs}}}
\label{reduced-redshift}
\end{equation}
can be calculated. An advantage of this method is that it provides simple, visible insight into the effect of a varying $\mu$ by plotting
the reduced redshifts $\zeta_i$ as a function of the sensitivity coefficient $K_i$ for each line. In view of the relation
\begin{equation}
  \zeta_i =  \frac{\Delta\mu}{\mu}K_i,
\label{reduced-method}
\end{equation}
the slope of the plot directly represents the $\Delta\mu/\mu$ value, while outliers in the data points are easily detected.
However, the line-by-line method is only suited for spectra containing isolated, singular absorption features of H$_2$, as is the case for system Q0347$-$383 \cite{Wendt2011,Wendt2012}. Even then, an absorption profile that appears singular might in fact be a sum of multiple narrowly separated velocity components, and the only robust means for testing that possibility is to fit all H$_2$ transitions simultaneously, as in the `comprehensive fitting' method discussed below and used for most analyses to date.

\begin{figure}
\centering
\includegraphics[width=0.96\columnwidth]{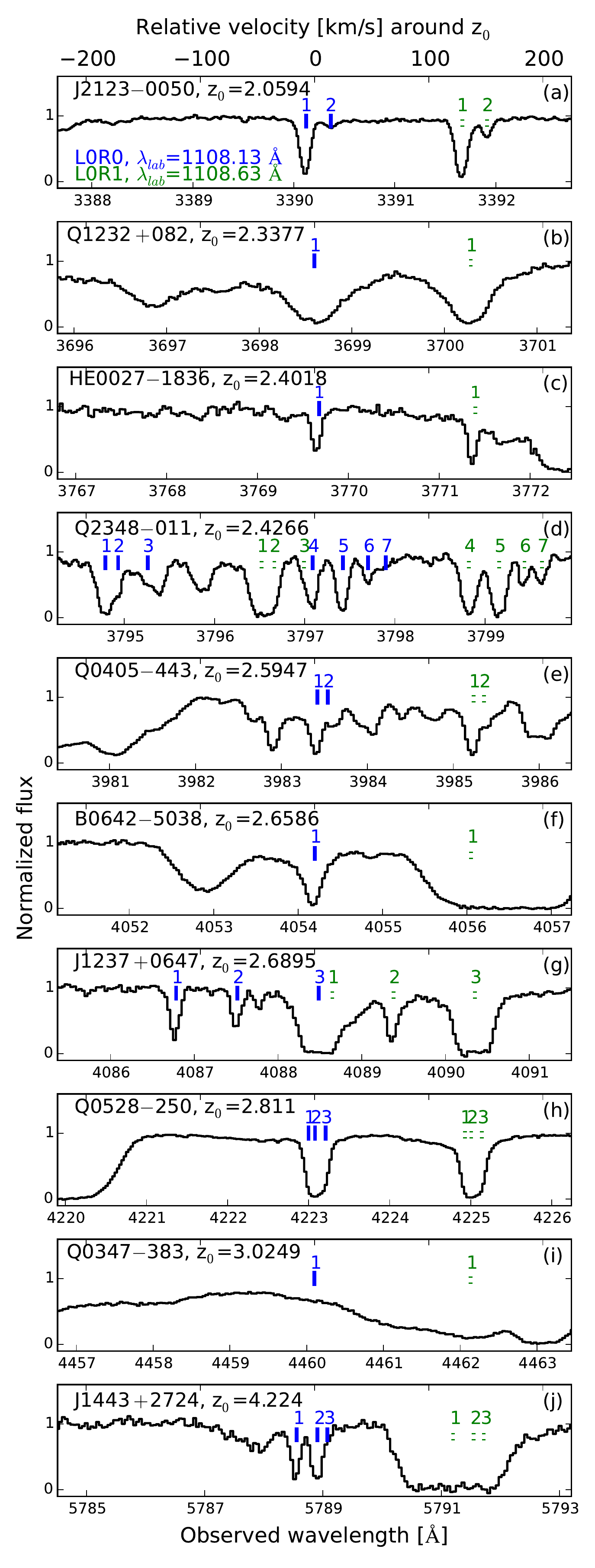}
\caption{The H$_2$ transitions L0R0 indicated with solid (blue) markers and L0R1 indicated with dashed (green) markers seen at different redshifts toward ten quasars. Transitions are imprinted in a number of velocity features as indicated. For further details see main text.
}
\label{comparison}
\end{figure}

Complex velocity profiles can be disentangled using a different fitting method which is known as the comprehensive fitting method. Within this approach, all available transitions are fitted simultaneously and solved for a single redshift for each identifiable absorbing H$_2$ velocity component, e.g.~see~\citet{King2008}, \citet{Malec2010}, and \citet{Bagdonaite2014a}. Spectral regions where H$_2$ profiles are partially overlapped by the Lyman-$\alpha$ forest, metal absorption lines, or other H$_2$ transitions  can be handled in this way as well, leading to an increased amount of information extracted from a single quasar spectrum. This is particularly the case if the H$_2$ profile is broad as demonstrated in the analysis of Q2348$-$011; see Fig.~\ref{comparison} and \citet{Bagdonaite2012}.

The software package VPFIT\footnote{\url{http://www.ast.cam.ac.uk/~rfc/vpfit.html}} is commonly used to implement the comprehensive fitting analysis. VPFIT is a non-linear least-squares $\chi^2$-minimization program dedicated specifically for analyzing quasar spectra composed of complex overlaps of multiple Voigt profiles.
The analysis starts by selecting a set of spectral regions where the H$_2$ transitions can be visually discerned from the Lyman-$\alpha$ forest. In each region, a different effective continuum is set by the broad forest lines but an H$_2$ velocity structure is seen repetitively and, therefore, fitting it requires fewer free parameters compared to the case of unrelated lines. For a single H$_2$ velocity component, all the transitions share the same redshift $z_{\rm{abs}}$, width $b$, and the same column density $N_J$ for transitions from the same rotational level $J$. By invoking such a parameter linking, one is able to minimize the total number of free parameters for H$_2$ and to gain information from multiple regions simultaneously. The laboratory wavelengths $\lambda_i^0$, the oscillator strengths $f_i$, and the damping parameters $\Gamma_i$ are the fixed parameters of the fit which are known from laboratory measurements or molecular physics calculations (see section~\ref{H2-spec}).

The comprehensive fitting method permits fitting complex velocity structures. The necessity to add more components to an H$_2$ profile is evaluated by inspecting the residuals of fitted regions individually and combined. The latter approach results in a 'composite residual spectrum' \cite{Malec2010} where normalized residuals of multiple regions are aligned in velocity or redshift space and averaged together. Any consistent underfitting is seen more clearly in the composite residual spectrum than in individual residuals. Naturally, various goodness-of-fit indicators (e.g.~a reduced $\chi^2$) are also used to justify addition of more components. Once a satisfactory fit is achieved, a final free parameter, $\Delta\mu/\mu$, is introduced which permits relative shifts of the H$_2$ transitions according to their $K_i$ coefficients. A single $\Delta\mu/\mu$ value is then extracted from the multiple fitted transitions of H$_2$.

\subsection{Known H$_2$ absorption systems at moderate-to-high redshift}
\label{known-systems}

In Table~\ref{Table-quasars} a listing is given of moderate-to-high redshift H$_2$ absorption systems as obtained from the literature. The majority of these quasars were observed with UVES-VLT, while system J2123$-$005 was observed both at Keck and VLT, J0812$+$3208 was observed at Keck, and the lower redshift systems HE0515$-$441 and Q1331$+$170 were observed with the Space Telescope Imaging Spectrograph aboard the Hubble Space Telescope.  Some of the lower redshift systems, Q1331$+$170 and Q1441$+$014, were observed at VLT specifically for the electronic CO absorptions. The Table is restricted to H$_2$ absorption systems with $z>1$. Besides those detections, some ten lower redshift (0.05 $\lesssim z \lesssim$ 0.7)~molecular hydrogen absorbers have been discovered as well in the archival Hubble Space Telescope/Cosmic Origins Spectrograph spectra (at log$N[$H$_2/$cm$^{-2}] > 14.4$~\cite{Muzahid2015}).

Even though H$_2$ is the most abundant molecule in the universe, its detections are rather scarce outside the Local Group of galaxies.
As a rule, molecular hydrogen absorption is associated with DLAs which are huge reservoirs of neutral hydrogen gas with log$N$[H$\textsc{i}$/cm$^{-2}]\geq20.3$. In a recent study of 86 medium-to-high resolution quasar spectra that contain DLAs, only a 1\,\% ($<$6\,\% at 95\,\% confidence) detection rate of H$_2$ was reported for log$N$[H$_2/$cm$^{-2}]>17.5$~\cite{Jorgenson2014}.

The most recent quasar census from the Sloan Digital Sky Survey (SDSS) contains over 80\,000 spectra of high-redshift quasars \cite{Paris2014} that were searched for H$_2$ absorption. The intermediate spectral resolution of $\sim 2000$, the usually low signal-to-noise ratio, and the presence of the Lyman-$\alpha$ forest overlapping the H$_2$ region complicates the identification of H$_2$ absorption systems. However, \citet{Balashev2014} devised an automated search and identification procedure, including a quantitative assessment of the H$_2$ detection probability as well as a false detection estimate.
This procedure yielded the identification of 23 highly likely H$_2$ absorption systems at column densities log$N$[H$_2/$cm$^{-2}]>18.8$  in the redshift interval $z=2.47-3.68$. As an example, the system J2347$-$005 at $z=2.59$ displays a robust an unambiguously assignable H$_2$ absorption spectrum. For conciseness this sample of 23 additional systems is not copied into Table~\ref{Table-quasars}, but it should definitely be regarded as part of the pool of possible targets for future investigations of $\mu$-variation based on the H$_2$ method.

\begin{table*}
\caption{List of known moderate-to-high redshift H$_2$ absorption systems with some relevant parameters. Bessel $R$ magnitude taken from the SuperCOSMOS Sky Survey~\cite{Hambly2001}. The ten systems analyzed so far for $\mu$-variation are offset at the top. The column densities $N$(H$_2$), $N$(HD), $N$(CO) and $N$(H\textsc{i}) are given on a $\log_{\rm{10}}$ scale in cm$^{-2}$. Some systems were investigated for CO only, and hence no column densities determined for H and H$_2$. It is noted that the 23 systems listed in Table I of \citet{Balashev2014} should also be regarded as very likely H$_2$ absorption systems.}
\label{Table-quasars} 
\begin{tabular}{p{2.5cm}p{0.8cm}p{0.8cm}p{1.8cm}p{1.9cm}p{1.5cm}p{1.2cm}p{1.2cm}p{1.3cm}p{1.3cm}p{1.9cm}}
\hline \hline
Quasar & $z_{\rm{abs}}$ & $z_{\rm{em}}$     & RA(J2000)  & Decl.(J2000) & $N$(H$_2$) & $N$(HD) & $N$(CO) & $N$(H) & $R_{\rm{mag}}$ & Refs. \\ \hline
\textbf{HE0027$-$1836}& $2.42$  & $2.55$ & 00:30:23.62 & $-$18:19:56.0 & $17.3$ &        & & $21.7$ & $17.37$  &   [1,2]  \\
\textbf{Q0347$-$383} & $3.02$  & $3.21$ & 03:49:43.64 & $-$38:10:30.6 & $14.5$ &        & & $20.6$ &$17.48$   &    [3-8] \\
\textbf{Q0405$-$443} & $2.59$  & $3.00$ & 04:07:18.08 & $-$44:10:13.9 & $18.2$ &        & & $20.9$ & $17.34$  &    [3-5,8]   \\
\textbf{Q0528$-$250} & $2.81$  & $2.81$ & 05:30:07.95 & $-$25:03:29.7 & $18.2$ & $13.3$ & & $21.1$ & $17.37$  &    [5,9]  \\
\textbf{B0642$-$5038} & $2.66$  & $3.09$ & 06:43:26.99 & $-$50:41:12.7 & $18.4$ &        & & $21.0$ & $18.06$  &   [10-12] \\
\textbf{Q1232$+$082} & $2.34$  & $2.57$ & 12:34:37.58 & $+$07:58:43.6    & $19.7$ & $15.5$ & & $20.9$ & $18.40$  & [13,14] \\
\textbf{J1237$+$064} & $2.69$  & $2.78$ & 12:37:14.60 & $+$06:47:59.5 & $19.2$& $14.5$& $14.2$ & $20.0$ &  $18.21$ & [15] \\
\textbf{J1443$+$2724} & $4.22$  & $4.42$ & 14:43:31.18 & $+$27:24:36.4 & $18.3$ &        & & $21.0$ & $18.81$  &   [16,17]   \\
\textbf{J2123$-$0050} & $2.06$  & $2.26$ & 21:23:29.46 & $-$00:50:52.9 & $17.6$ & $13.8$ & & $19.2$ & $15.83$  &   [18,19] \\
\textbf{Q2348$-$011} & $2.42$  & $3.02$ & 23:50:57.87 &  $-$00:52:09.9 & $18.4$ &        & & $20.5$  & $18.31$  &  [20-22] \\
\\
\textbf{J0000$+$0048} & $2.52$  &3.03  & 00:00:15.17 & $+$00:48:33.29  &  &         & $15$ &    & $19.2$   &   [23] \\
\textbf{Q0013$-$004} & $1.97$  & $2.09$ & 00:16:02.40 & $-$00:12:25.0  & $18.9$ &        & & $20.8$ & $17.89$  &   [24]  \\
\textbf{Q0201$+$113} & 3.39 & 3.61 & 02:03:46.66 & $+$11:34:45.4 		& $\leq$16.4 &    & & 21.3  & 19.41    &   [25]  \\
\textbf{HE0515$-$441} & 1.15 & 1.71 & 05:17:07.63 & $-$44:10:55.5 &  &  &  & 19.9 & 14.0 & [26] \\
\textbf{Q0551$-$366} & $1.96$  & $2.32$ & 05:52:46.18 & $-$36:37:27.5  & $17.4$ &        & & $20.5$ & $17.79$  &   [27]  \\
\textbf{J0812$+$3208}& $2.63$  & $2.70$ & 08:12:40.68  & $+$32:08:08.6 & $19.9$ & $15.4$ & & $21.4$ & $17.88$ &    [28-30]  \\
\textbf{J0816$+$1446}& $3.29$ & 3.85 & 08:16:34.39 & $+$14:46:12.5 & 18.66 &       & & 22.0   &   $19.20$ &    [31]  \\
\textbf{Q0841$+$129} & $2.37$  & $2.48$ & 08:44:24.24 & $+$12:45:46.5    & $14.5$ &      & & $20.6$ & $17.64$ &    [32]   \\
\textbf{J0857$+$1855} & 1.72 & 1.89 & 08:57:26.79 & $+$18:55:24.4        &        &      & 13.5  &  & $17.32$&     [33]  \\
\textbf{J0918$+$1636} & 2.58   & 3.07   & 09:18:26.1  & $+$16:36:09      & $\leq19.0$ &  & & $21.0$ & $19.49$   &  [34] \\
\textbf{J1047$+$2057} & 1.77 & 2.01 & 10:47:05.8  & $+$20:57:34      &        &        & $14.7$  & & $19.96$ & [33] \\
\textbf{Q1331$+$170} & $1.78$  & $2.08$ & 13:33:35.81 & $+$16:49:03.7    & $19.7$ & $14.8$ & & $21.2$ & $16.26$  & [28,35] \\
\textbf{J1337$+$3152} & $3.17$  & $3.17$ & 13:37:24.69 & $+$31:52:54.6   & $14.1$ &        & & $21.4$ & $18.08$   & [36] \\
\textbf{J1439$+$1117} & $2.42$  & $2.58$ & 14:39:12.04 & $+$11:17:40.5   & $19.4$ & $14.9$ & 13.9 & $20.1$ & $18.07$  & [37,38] \\
\textbf{Q1444$+$014} & $2.08$  & $2.21$ & 14:46:53.04 & $+$01:13:56.0    & $18.3$ &        & & $20.1$ & $18.10$  & [39] \\
\textbf{J1456$+$1609} & 3.35 &3.68 & 14:56:46.481 & $+$16:09:39.31  &   $17.1$ &        & &  $21.7$  & $19.05$ & [40] \\
\textbf{J1604$+$2203} & 1.64 & 1.98 & 16:04:57.49 & $+$22:03:00.7        &        &        & $14.6$ &   & $19.09$  & [41] \\
\textbf{J1705$+$3543} & 2.04 & 2.02$^a$  & 17:05:42.91 & $+$35:43:40.3   &  & & $14.1$& & 19.42 & [33] \\
\textbf{Q2100$-$0641} & 3.09 & 3.14 & 21:00:25.029 & $-$06:41:45.99    & $18.8$ &        & &  $21.05$  & $17.52$ & [42] \\
\textbf{J2140$-$0321} & 2.34 & 2.48 & 21:40:43.016 &    $-$03:21:39.29    &   $20.1$ &        & &  $22.4$  & $18.93$ & [23,42] \\
\textbf{Q2318$-$111} & $1.99$  & $2.56$ & 23:21:28.69 & $-$10:51:22.5    & $15.5$ &        & & $20.7$ & $17.68$  & [1] \\
\textbf{J2340$-$0053} & 2.05 & 2.09 & 23:40:23.67 & $-$00:53:27.00  &   $18.2$ &        & & $20.35$  & $17.66$   & [30,42] \\
\textbf{Q2343$+$125} & $2.43$  & $2.52$ & 23:46:25.42 & $+$12:47:43.9    & $13.7$ &        & & $20.4$ & $20.18$  & [43,44] \\
\hline
\multicolumn{11}{l}{{$^a$ $z_{\rm{em}}$ reported by~\citet{Hewett2010} is smaller than $z_{\rm{abs}}$ from~\citet{Noterdaeme2011}.}}\\
\multicolumn{11}{l}{References: [1]~\citet{Noterdaeme2007b}; [2]~\citet{Rahmani2013}; [3]~\citet{Ivanchik2005}; [4]~\citet{Reinhold2006};}\\ \multicolumn{11}{l}{[5]~\citet{King2008}; [6]~\citet{Wendt2011}; [7]~\citet{Wendt2012}; [8]~\citet{Thompson2009};} \\
\multicolumn{11}{l}{[9]~\citet{King2011}; [10]~\citet{Noterdaeme2008}; [11]~\citet{Bagdonaite2014a}; [12]~\citet{Albornoz2014}; } \\
\multicolumn{11}{l}{[13]~\citet{Varshalovich2001}; [14]~\citet{Ivanchik2010}; [15]~\citet{Noterdaeme2010}; [16]~\citet{Ledoux2006b};} \\
\multicolumn{11}{l}{[17]~\citet{Bagdonaite2015}; [18]~\citet{Malec2010}; [19]~\citet{Weerdenburg2011}; [20]~\citet{Ledoux2006}; } \\
\multicolumn{11}{l}{[21]~\citet{Noterdaeme2007}; [22]~\citet{Bagdonaite2012}; [23]~\citet{Noterdaeme2015b}; [24]~\citet{Petitjean2002}; } \\
\multicolumn{11}{l}{[25]~\citet{Srianand2012}; [26]~\citet{Reimers2003}; [27]~\citet{Ledoux2002}; [28]~\citet{Tumlinson2010};  } \\
\multicolumn{11}{l}{[29]~\citet{Balashev2010}; [30]~\citet{Jorgenson2010}; [31]~\citet{Guimares2012}; [32]~\citet{Petitjean2000}; } \\
\multicolumn{11}{l}{[33]~\citet{Noterdaeme2011}; [34]~\citet{Fynbo2011}; [35]~\citet{Cui2005}; [36]~\citet{Srianand2010}; } \\
\multicolumn{11}{l}{[37]~\citet{Noterdaeme2008b}; [38]~\citet{Srianand2008}; [39]~\citet{Ledoux2003}; [40]~\citet{Noterdaeme2015}; } \\
\multicolumn{11}{l}{[41]~\citet{Noterdaeme2009}; [42]~\citet{Balashev2015}; [43]~\citet{Petitjean2006}; [44]~\citet{Dessauges2004}.} \\
\end{tabular}
\end{table*}

The 33 systems listed in Table~\ref{Table-quasars} added with the 23 very likely additional systems identified by~\citet{Balashev2014} makes up a currently available database of 56 H$_2$ known absorption systems that may be used for searches of $\mu$-variation.
However, not every H$_2$ detection proves to be equally useful in the analysis of $\mu$ variation -- some can be discarded because observational requirements are currently too challenging or particular absorbing systems have unsuitable intrinsic properties.
An ideal system should have:
\begin{itemize}
	\item A bright background source so that a high-resolution spectrum with a signal-to-noise ratio in the continuum of about 50 per 2.5\,km/s pixel could be obtained within, e.g.~15\,--\,20 hours at an 8--10\,m class optical telescope. This can be achieved for background quasars with $R_{\rm{mag}} = 17 - 18$.
	\item A column density of H$_2$ in the range between log$N$[H$_2$/cm$^{-2}]\sim14$ and 18. A column density outside this range would either yield a small number of detectable H$_2$ transitions or a high number of saturated transitions; neither of these situations is desirable since the precision at which $\Delta\mu/\mu$ can be measured depends on the number of strong but unsaturated transitions.
A column density as high as log$N$[H$_2$/cm$^{-2}] \sim 18 - 19$ however has the advantage that H$_2$ transitions from high-$J$ states in combination with larger numbers of HD and CO lines would become observable at high signal-to-noise.
	\item An absorption redshift of at least $z = 2$ or larger to assure that a sufficient number of lines (typically $>40$) will shift beyond the atmospheric cut-off at 3000 \AA.
    \item Absorption profiles of H$_2$ with simple substructure. While fitting multiple Voigt components to an H$_2$ absorption profile is feasible and can be justified, a simpler absorption substructure is preferred because it simplifies the analysis and tests for systematic errors.
\end{itemize}

A system that obeys these requirements would yield a $\Delta\mu/\mu$ constraint with a precision of several parts per million. Table~\ref{Table-quasars} contains a list of the 10 best H$_2$ systems that have been already analyzed for $\mu$ variation, and 23 additional systems whose properties seem less suitable for a varying-$\mu$ analysis, although not all have been investigated in sufficient detail. Relevant properties of all systems, such as absorption and emission redshifts, position on the sky, the known column densities for H$_2$, deuterated molecular hydrogen, HD, carbon monoxide, CO, and neutral atomic hydrogen, H, and the magnitude are provided in the Table.
With similar $K_i$ sensitivities as those of H$_2$, the rovibronic transitions of HD~\cite{Ivanov2010} and CO~\cite{Salumbides2012} provide a way to independently cross check $\Delta\mu/\mu$ constraints from H$_2$. However, the column densities of HD and CO are usually $\sim$10$^{5}$ times smaller than $N$(H$_2$), leading to a much smaller number of detections and fewer transitions in case of detection.

\subsection{Constraints from individual quasars}
\label{individual}

A comparison between the ten quasar absorption spectra, for which the H$_2$ absorption spectrum is analyzed in detail, is displayed in Fig.~\ref{comparison}. In each case the wavelength region of two H$_2$ absorption lines is covered, the L0R0 line at a rest wavelength of 1108.13 \AA\ and the L0R1 line at 1108.63 \AA, plotted on a wavelength scale and corresponding velocity scale (in km/s).
The spectra illustrate typical characteristics of high redshift H$_2$ absorption. First of all this figure exemplifies that the H$_2$ spectra for different absorbers fall in different wavelength ranges. The L0R0 and L0R1 lines, at the red side of the H$_2$ absorption spectrum, are detected at 3390 \AA\ in the ultraviolet for the lowest redshift system J2123$-$005, to 5789 \AA\ in the yellow range for the highest redshift system in the sample. The velocity structure varies strongly from one absorber to the next, displaying single velocity features as in B0642$-$5038 and HE0027$-$1836, to three features as in Q0528$-$250 and J1443$+$2724, to an amount of seven clearly distinguishable features in the case of Q2348$-$011. In more detailed analyses of the spectra additional underlying velocity components are revealed. The widths for both L0R0 and L0R1 lines in Q1232$+$082 are broadened due to saturation of the absorption in view of the large column density. Other lines in this sight line exhibit a width of 4.5~km/s. The broad absorption lines due to the Lyman-$\alpha$ forest are random, with two strong H${\textsc{i}}$ lines appearing in the displayed wavelength interval toward B0642$-$5038, one of them overlapping the L0R1 line. Similarly in the spectrum of J1443$+$2724 all three velocity features of the L0R1 line are hidden by a Lyman-$\alpha$ forest line. For Q0347$-$383 both L0R0 and L0R1 lines are covered by broad forest features. Here it is pointed out that the number-density of Lyman-$\alpha$ forest features increases with increasing redshift~\cite{Meiksin2009} making the analysis of absorption systems at higher $z$ gradually more difficult.

\begin{figure*}
\centering
\includegraphics[scale=0.95]{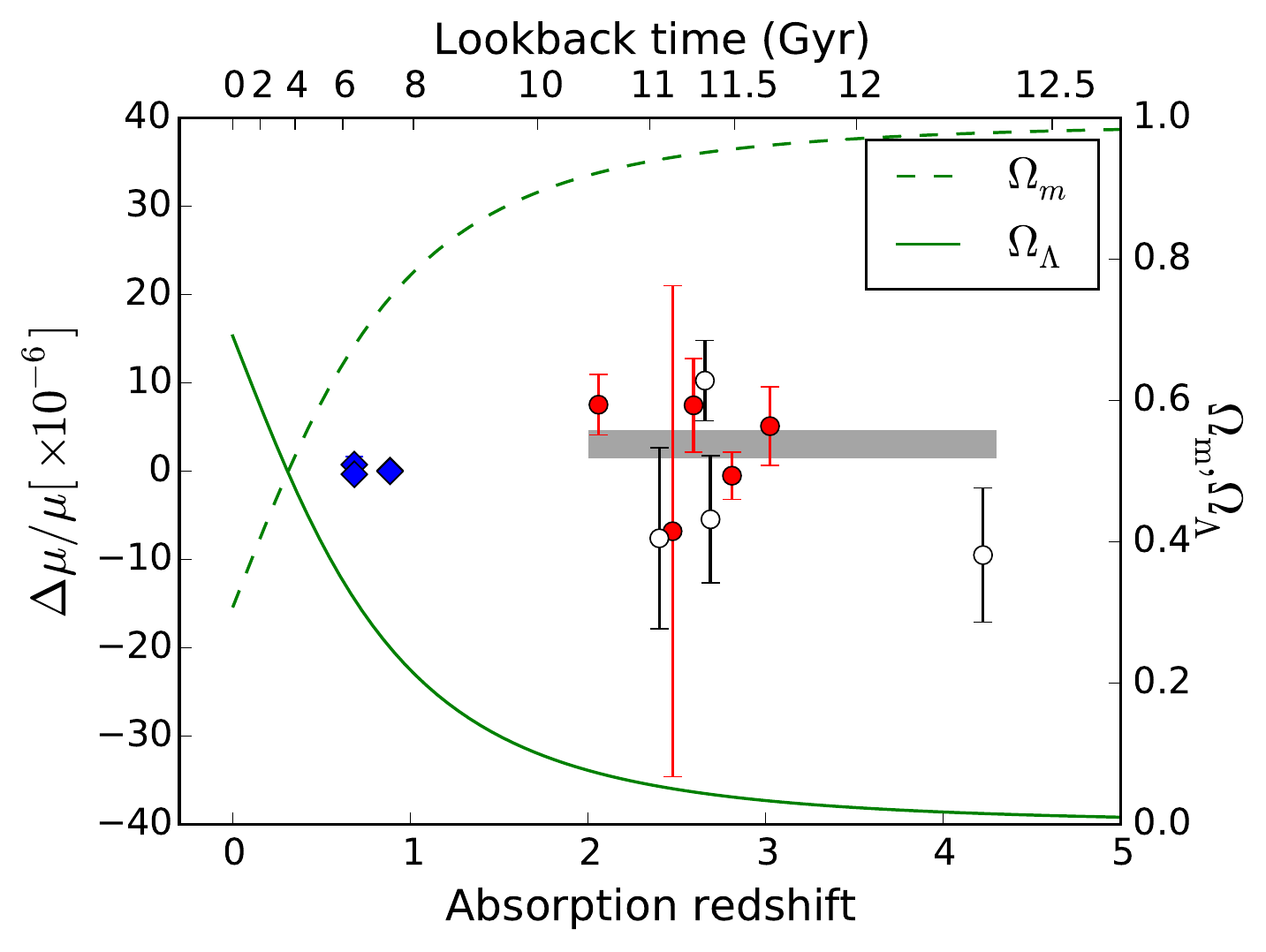}
\caption{Values for $\Delta\mu/\mu$ obtained for nine H$_2$ absorption systems analyzed as a function of redshift and look-back time plotted alongside with evolution of cosmological parameters for $\Omega_m$ (dashed line) and $\Omega_{\Lambda}$ (full line), referred to the right-hand vertical axis. These are the dark matter and dark energy densities, respectively, relative to the critical density. Open circles refer to data where a correction for long-lange wavelength distortions is included; for the filled (red) circles such correction has not been applied. Uncertainties for individual points are indicated at the 1$\sigma$ level by vertical line extensions to points. The result for system Q1232$+$082 is not plotted in view of its large uncertainty. The grey bar represents the average for $\Delta\mu/\mu$ with $\pm 1\sigma$ uncertainty limits from the 10 H$_2$ systems analyzed. For comparison results from radio astronomical observations at lower redshift ($z<1$) are plotted as well (diamonds - blue).}
\label{mu-plot-overview}
\end{figure*}

Details on the ten H$_2$ absorption systems for which a value for $\Delta\mu/\mu$ was deduced are as follows:
\begin{itemize}
	\item {\bf HE0027$-$1836}.
        This system at $z=2.40$ was observed over three years at VLT and an analysis involving calibrations with attached ThAr spectra as well as asteroid spectra, providing a correction for long-range wavelength distortions, yields a constraint of $\Delta\mu/\mu = (-7.6 \pm 8.1_{\rm{stat}} \pm 6.3_{\rm{sys}})\times 10^{-6}$~\cite{Rahmani2013}, which corresponds to $\Delta\mu/\mu = (-7.6 \pm 10.2)\times 10^{-6}$.

	\item {\bf Q0347$-$383}.
        First rounds of analysis for this system at $z=3.02$ were pursued by \citet{Ivanchik2002,Ivanchik2005} and by \citet{Reinhold2006} based on a line-by-line analysis method, in part relying on less accurate laboratory reference data from classical spectroscopy. Subsequently the comprehensive fitting method was employed by \citet{King2008} yielding $\Delta\mu/\mu = (8.5 \pm 7.4)\times 10^{-6}$. Advanced line-by-line re-analysis studies of the existing data for this object yielded $\Delta\mu/\mu = (-28 \pm 16)\times 10^{-6}$ \cite{Thompson2009}, while renewed observations yielded $\Delta\mu/\mu = (15 \pm 9_{\rm{stat}} \pm 6_{\rm{sys}})\times 10^{-6}$~\cite{Wendt2011} and  $\Delta\mu/\mu = (4.3 \pm 7.2)\times 10^{-6}$ \cite{Wendt2012}. A weighted average over the latter four most accurate results yields $\Delta\mu/\mu = (5.1 \pm 4.5)\times 10^{-6}$.

	\item {\bf Q0405$-$443}.
        Similarly as for the previous item some early studies were performed on this system at $z=2.59$ \cite{Ivanchik2005,Reinhold2006}.
        A study using the comprehensive fitting method \cite{King2008} yielded $\Delta\mu/\mu = (10.1 \pm 6.2)\times 10^{-6}$ and a line-by-line fitting study \cite{Thompson2009} yielded $\Delta\mu/\mu = (0.6 \pm 10)\times 10^{-6}$. As an average value we adopt $\Delta\mu/\mu = (7.5 \pm 5.3)\times 10^{-6}$. Although this system exhibits two absorption features (see Fig.~\ref{comparison}) the weaker one was left out in all $\mu$-variation analyses performed so far.

    \item {\bf Q0528$-$250}. This system at $z=2.81$ was subject to the first constraint on a varying $\mu$ with observations from the MMT by~\citet{Foltz1988} yielding $\Delta\mu/\mu < 2 \times 10^{-4}$. Further early stage observations were performed with the Keck telescope by \citet{Cowie1995}. Based on observations at VLT \cite{Ledoux2003} an accurate analysis was performed using the comprehensive fitting method
        yielding $\Delta\mu/\mu = (-1.4 \pm 3.9)\times 10^{-6}$ \cite{King2008}. Later renewed VLT observations were conducted with ThAr attached calibrations yielding $\Delta\mu/\mu = (0.3 \pm 3.2_{\rm{stat}} \pm 1.9_{\rm{sys}})\times 10^{-6}$ \cite{King2011}. As an average value we adopt $\Delta\mu/\mu = (-0.5 \pm 2.7)\times 10^{-6}$.

 	\item {\bf B0642$-$5038}.
        This system at $z=2.66$ at the most southern declination (dec = $-50^o$) exhibits a single H$_2$ absorption feature, that was analyzed in a line-by-line analysis, yielding $\Delta\mu/\mu = (7.4 \pm 4.3_{\rm{stat}} \pm 5.1_{\rm{sys}})\times 10^{-6}$~\cite{Albornoz2014} and in a comprehensive fitting analysis, yielding $\Delta\mu/\mu = (12.7 \pm 4.5_{\rm{stat}} \pm 4.2_{\rm{sys}})\times 10^{-6}$~\cite{Bagdonaite2014a}. The latter study includes a correction for long-range wavelength distortions by comparing to asteroid and solar twin spectra. From these studies we adopt an averaged value $\Delta\mu/\mu = (10.3 \pm 4.6)\times 10^{-6}$.

    \item {\bf Q1232$+$082}. This system at $z=2.34$ exhibits strong absorption in a single feature with strongly saturated lines for low-$J$ entries. From over 50 absorption lines a selection was made of 12 isolated, unsaturated and unblended lines that were compared in a line-by-line analysis with two laboratory wavelength sets (now outdated) to yield $\Delta\mu/\mu = (14.4 \pm 11.4)\times 10^{-5}$ and $\Delta\mu/\mu = (13.2 \pm 7.4)\times 10^{-5}$~\cite{Ivanchik2002}, which averages to $\Delta\mu/\mu = (140 \pm 60)\times 10^{-6}$. This constraint is much less tight than from the other studies in view of the limited data set of unsaturated lines. It is therefore not included in Fig.~\ref{mu-plot-overview}, although the result is included in the calculation of the total average.

    \item {\bf J1237$+$064}. This system at $z=2.69$ exhibits three distinct H$_2$ absorption features, one of which is strongly saturated for the low-$J$ components. An analysis of over 100 lines of H$_2$ and HD yields a value of $\Delta\mu/\mu = (-5.4 \pm 6.3_{\rm{stat}} \pm 3.5_{\rm{sys}})\times 10^{-6}$~\cite{Dapra2015}. This value represents a result after invoking a long-range distortion-correction analysis. Combining the uncertainties we adopt the value $\Delta\mu/\mu = (-5.4 \pm 7.2)\times 10^{-6}$ in our analysis.

    \item {\bf J1443$+$2724}. This system at the highest redshift ($z=4.22$) and the most northern (dec = $27^o$) probed was observed with UVES-VLT. For the $\mu$-variation analysis archival data from 2004 \cite{Ledoux2006} were combined with data recorded in 2013 yielding a constraint from an analysis also addressing the problem of long-range wavelength distortions in the ThAr calibrations:
        $\Delta\mu/\mu = (-9.5 \pm 5.4_{\rm{stat}} \pm 5.3_{\rm{sys}})\times 10^{-6}$~\cite{Bagdonaite2015}, which corresponds to $\Delta\mu/\mu = (-9.5 \pm 7.5)\times 10^{-6}$.

    \item {\bf J2123$-$005}. This system at $z=2.05$ was observed with the highest resolution ever for any H$_2$ absorption system with the HIRES-spectrometer on the Keck telescope using a slit width of 0.3" delivering a resolving power of 110,000. A comprehensive fitting analysis yielded a value of $\Delta\mu/\mu = (5.6 \pm 5.5_{\rm{stat}} \pm 2.9_{\rm{sys}})\times 10^{-6}$~\cite{Malec2010}.
        Independently a spectrum was observed, in visitor mode, using the UVES spectrometer at the VLT, delivering a value of
        $\Delta\mu/\mu = (8.5 \pm 3.6_{\rm{stat}} \pm 2.2_{\rm{sys}})\times 10^{-6}$~\cite{Weerdenburg2011}. Averaging over these independent results yields  $\Delta\mu/\mu = (7.6 \pm 3.5)\times 10^{-6}$.

    \item {\bf Q2348$-$011}.
        This system at $z=2.43$ exhibits a record complex velocity structure with seven distinct H$_2$ absorption features, and some additional underlying substructure. Nevertheless a comprehensive fitting could be applied to this system yielding a constraint of $\Delta\mu/\mu = (-6.8 \pm 27.8)\times 10^{-6}$~\cite{Bagdonaite2012}.

\end{itemize}

\subsection{Current high-redshift constraint on $\Delta\mu/\mu$}
\label{current}

For the ten systems analyzed, results for $\Delta\mu/\mu$ were determined by various authors. Statistical and systematic uncertainties were combined here and the results from the various analyses were averaged to yield a single value of $\Delta\mu/\mu$ with an associated 1-$\sigma$ uncertainty for each system.
These results for the 10 H$_2$ quasar systems are plotted in Fig.~\ref{mu-plot-overview} (except for Q1232+082) along with the more precise results from radio astronomical observations at redshifts $z=0.68$ and $z=0.89$. The final result is obtained by taking a weighted average over the outcome for all 10 systems in the interval $z=2.0-4.2$, yielding $\Delta\mu/\mu = (3.1 \pm 1.6)\times 10^{-6}$. This result shows a larger proton--electron mass ratio in the past at a significance of 1.9-$\sigma$. We interpret this result as a null result bearing insufficient significance to call for new physics in terms of varying constants.

\begin{figure*}
\centering
\includegraphics[scale=0.95]{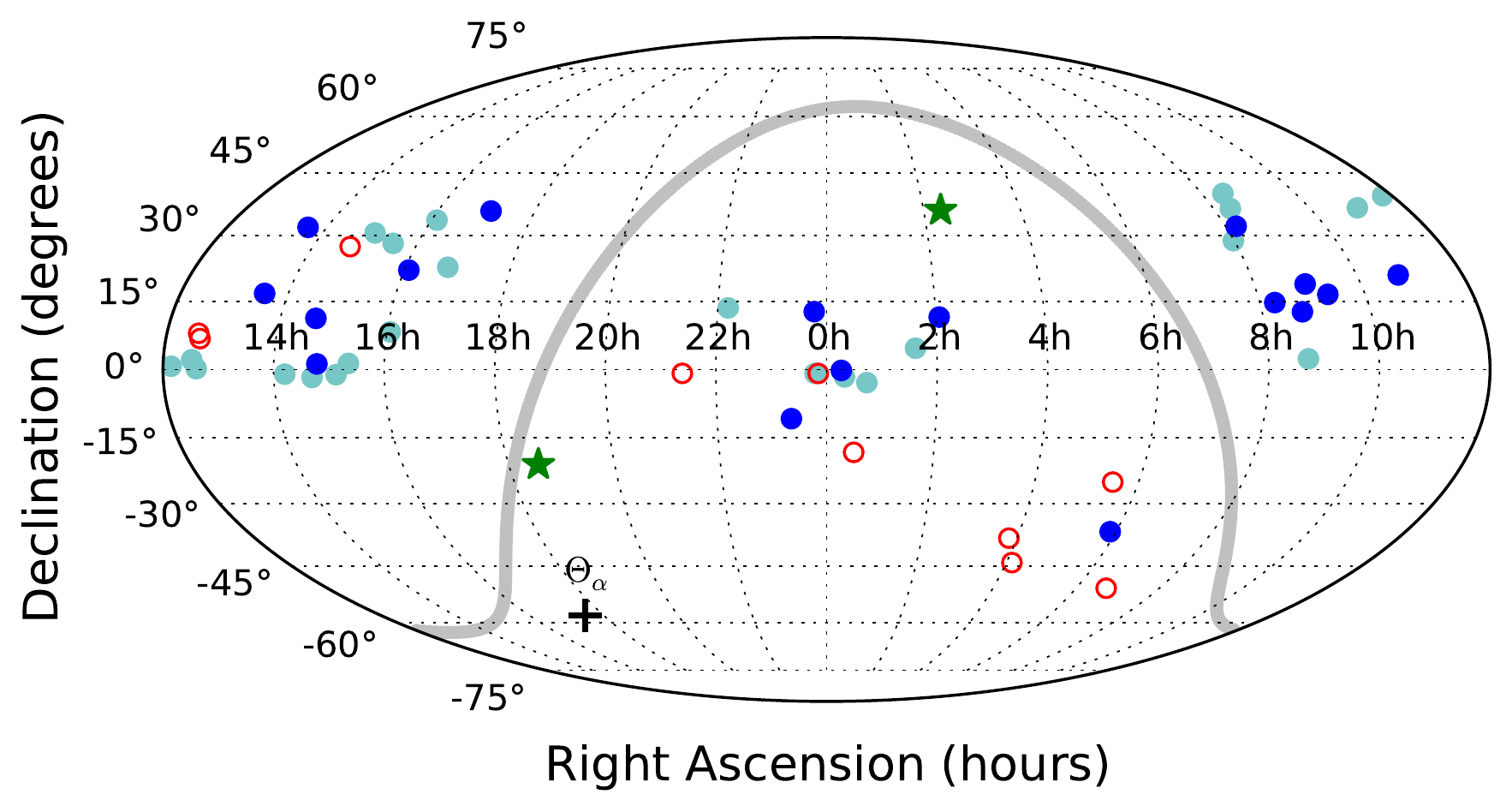}
\caption{Sky map in equatorial coordinates (J2000) showing currently known quasar sightlines containing molecular absorbers at intermediate-to-high redshifts. The open (red) points correspond to the 10 H$_2$ absorbers that have been analyzed for a variation of $\mu$, while the dark (blue) points mark the 23 H$_2$ targets that might be used in future analyses as listed in Table~\ref{Table-quasars}. The grey (light blue) points correspond to the additional sample of 23 H$_2$ absorption systems found through the Sloan Digital Sky Survey~\cite{Balashev2014}. The (green) stars indicate the sightlines toward PKS1830$-$211 and B0218$+$357 where $\mu$ variation was measured from CH$_3$OH~\cite{Bagdonaite2013a,Bagdonaite2013b} and NH$_3$~\cite{Henkel2009}, or only from NH$_3$~\cite{Murphy2008a,Kanekar2011}, respectively. The (+) sign represents the angle $\Theta_{\alpha}$ of the $\alpha$-dipole phenomenon~\cite{Webb2011}. The grey line indicates the Galactic plane.}
\label{skymap}
\end{figure*}

It is noted further that for some of the results from individual quasars the long-range wavelength distortions found recently in ESO-VLT spectra  were addressed, either by correcting for the distortion and/or as an increase of the systematic uncertainty. This was done for systems HE0027$-$1836~\cite{Rahmani2013}, B0642$-$5038~\cite{Bagdonaite2014a}, J1443$+$2724~\cite{Bagdonaite2015}, and J1237$+$064~\cite{Dapra2015}. For the other systems no such corrections were carried out, e.g.~for the reason that no super-calibration spectra are available or that at the time of analysis this systematic effect was not yet known, and the values adopted in Fig.~\ref{mu-plot-overview} are as published. So for some of the latter systems there may still exist the underlying systematic effect connected to long-range wavelength distortions of the calibration scale. From the distortion analyses carried out it follows that for most cases the distortion gives rise to an increase of $\Delta\mu/\mu$ by a few times $10^{-6}$, which is commensurate with the finding of a positive value of similar size. A re-analysis of possible wavelength distortions on some systems would be called for, but unfortunately solar-twin or asteroid spectra were not always recorded in exactly the time periods of the quasar observations. This open-ended calibration issue prompts the assignment of an increased systematic error to the combined result we presented above. Awaiting further detailed assessment, we conservatively estimate a null result of $|\Delta\mu/\mu| < 5 \times 10^{-6}$ (3-$\sigma$) for redshifts in the range $z=2.0-4.2$.

In Fig.~\ref{mu-plot-overview}  an explicit connection is made between varying constants over a range of redshifts and the evolution of matter density in the Universe, a connection also made in theoretical models predicting the temporal dependence of the values of the coupling constants. In view of the coupling of the additional dilaton fields to matter, constants should not drift in a dark-energy-dominated Universe~\cite{Barrow2002,Sandvik2002,Barrow2005}. All H$_2$ data are obtained for redshifts $z>2$, or look-back times in excess of 10 billion years, in the epoch of a matter-dominated Universe with $\Omega_m > 0.9$. According to prevailing theoretical models, if any variation of constants should occur, it should fall in this early stage of evolution. It must be concluded that the H$_2$ data of Fig.~\ref{mu-plot-overview} either do not comply with theoretical predictions of varying constants, or that the experimental findings are not sufficiently constraining to verify/falsify the predictions.

\subsection{Spatial effect on $\mu$}

Analysis of over 300 measurements of ionized metallic gas systems in quasar absorption spectra have revealed a possible spatial variation of the fine-structure constant $\alpha$. This spatial distribution effect was represented in terms of a dipole, as a first order approximation of a spatial effect. The statistical analysis of the comprehensive data set yields the dipole to fall along the axis given by coordinates $RA = 17.5 \pm 0.9$ hrs and dec =$ -58 \pm 9$ degrees at a 4.2-$\sigma$ significance level~\cite{Webb2011,King2012}. The direction of this dipole axis $\Theta_{\alpha}$ for the spatial effect on $\alpha$ is indicated in Fig.~\ref{skymap}. Recently it was argued by \citet{Whitmore2015} that the long-range wavelength distortions on the Th-Ar calibrations could be held responsible for at least part of the reported $\alpha$-dipole phenomenon. In this section we wish to explore a possible spatial effect on $\mu$ to be deduced from the analyzed hydrogen absorbers.

Fig.~\ref{skymap} graphically displays the spatial distribution of all 56 currently known H$_2$ absorption systems plotted in an angular coordinate system [$RA$, dec]. The red data points represent the 10 sightlines for which detailed H$_2$ analyses were performed, while the possible H$_2$ target systems are displayed as blue data points, dark blue for the 23 quasars listed in Table~\ref{Table-quasars}, and light blue for the 23 systems of~\citet{Balashev2014}. Inspection of this sky-map figure illustrates the remarkable feature that the 10 investigated H$_2$ absorbing systems all fall in a narrow band across the sky, perpendicular to the $\alpha$-dipole angle $\Theta_{\alpha}$. This feature of a biased data set dictates that a search for a spatial effect in a three-dimensional space is severely hampered.

Nevertheless attempts to uncover a possible spatial effect were pursued based on the data for $\Delta\mu/\mu$. Firstly an unconstrained fit was performed to extract an optimized dipole axis from the nine most precisely measured systems, Q1232$+$085 being too imprecise. This yields an axis at $RA = 4.9 \pm 4.8$ hrs and dec = $-66 \pm 30$ degrees and an amplitude of $(9.5 \pm 5.3) \times 10^{-6}$. The optimum values span a plane perpendicular to the dipole axis $\Theta_{\alpha}$, but the resulting values for the angles are ill defined. This outcome of the fitting procedure, which is likely to optimize for a spread of values in their plane of reference, may indeed be regarded as an artefact of the biased data set.

\begin{figure}
\centering
\includegraphics[width=1.0\columnwidth]{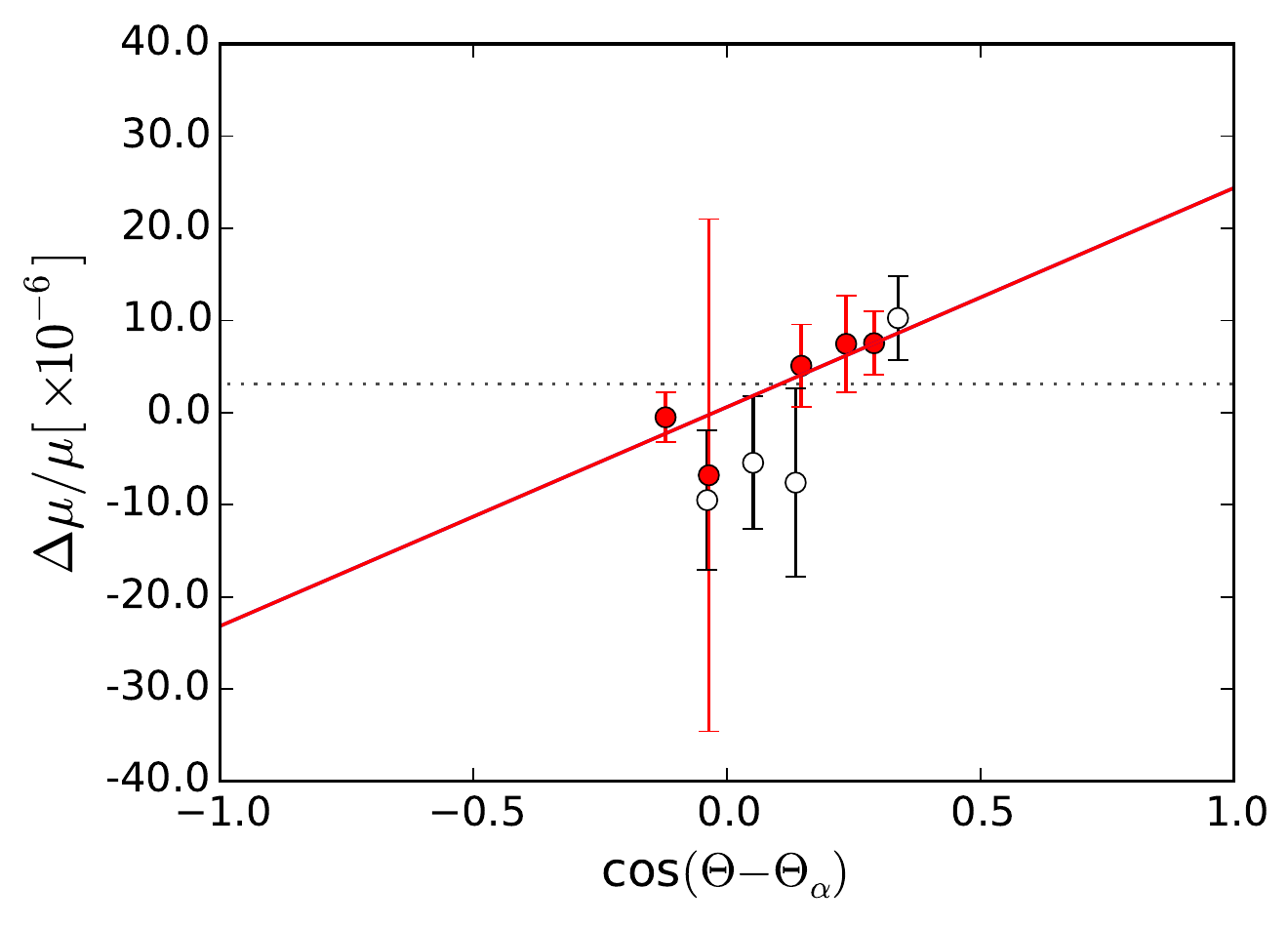}
\caption{Values of $\Delta\mu/\mu$ of the results for H$_2$ absorbers obtained so far plotted with their angular coordinates in terms of a projection angle $\Theta$ onto the dipole axis $\Theta_{\alpha}$ ($RA = 17.5$ hrs and dec =$ -58$ degrees) associated with the dipole found for $\Delta\alpha/\alpha$ \cite{Webb2011}. Again open circles represent values of $\Delta\mu/\mu$ that were corrected for possible long-range wavelength distortions, while filled (red) circles refer to uncorrected data points. The full line represents the result from a fit to Eq.~(\ref{fit-spatial}) with the slope equalling $A_{\Theta}$. See main text for several caveats in interpreting the significance of this fit.}
\label{Spatial-fig}
\end{figure}

Secondly an attempt was made to assess whether the $\Delta\mu/\mu$ measurements are consistent with the putative $\alpha$-dipole axis determined by \citet{Webb2011} at $\Theta_{\alpha}$ = [$RA=17.5$ hrs, dec = -58 degrees]. For this purpose the projection angles $\Theta_{\mu}$ of the H$_2$ absorbers were calculated with respect to the axis $\Theta_{\alpha}$ and plotted in Fig.~\ref{Spatial-fig}. This plot illustrates once more that the H$_2$ absorbers all lie at values of $\cos[\Theta_{\mu}-\Theta_{\alpha}] \approx 0$, hence $\Theta_{\mu} \perp \Theta_{\alpha}$.
Subsequently a fit was performed to the functional form
\begin{equation}
\Delta\mu/\mu = A_0 + A_{\Theta} \cos[\Theta_{\mu}-\Theta_{\alpha}],
\label{fit-spatial}
\end{equation}
yielding a dipole amplitude of $A_{\Theta} = (2.39 \pm 0.68) \times 10^{-5}$,  representing an effect at the 3-$\sigma$ significance level.
Fitting a possible linear offset returns $A_0= (0.6 \pm 1.4) \times 10^{-6}$ which makes a monopole contribution insignificant.
It is noted here that the biased data apparently exaggerates a trend along the perpendicular direction to $\Theta_{\alpha}$.
If anything, these analyses show the importance of investigating H$_2$ absorbing systems that fall outside the band of the 10 analyzed H$_2$ absorber systems in order to uncover possible spatial effects of $\mu$-variation.

\section{Perspectives}

The results on $\Delta\mu/\mu$ obtained from analysis of H$_2$ spectra from VLT and Keck observations, now yielding a 3-$\sigma$ constraint $|\Delta\mu/\mu| < 5\times 10^{-6}$, are not easy to improve with existing technologies.
The best known H$_2$ absorption systems, fulfilling the requirements for obtaining high-quality spectra, now numbering 10 (see Table~\ref{Table-quasars}), have been analyzed in detail.
There exist additional high-redshift systems with H$_2$ absorption (also listed in Table~\ref{Table-quasars}), but those exhibit less favorable parameters, i.e. either too low or too high molecular column densities, have too low redshifts for obtaining a full H$_2$ spectrum from Earth-based telescopes, or the back-ground quasar appears faint. Nevertheless additional observations with in-depth analyses will enlarge the database and increase the statistical significance of the constraint on $\Delta\mu/\mu$ in the redshift range $z=2.0-4.2$. First of all the systematic long-range wavelength distortions must be addressed by (i) observing solar-twin and asteroid spectra attached to quasar spectra and ThAr calibrations, and therewith correct for the distortions, and (ii) find and repair the causes of the distortions in the instruments.

Major improvements can be gained from technological advances, reducing the statistical and systematic uncertainties, which are almost equally limiting the accuracy at the level of few $10^{-6}$. The development of the next generation telescopes with larger collecting dishes, i.e.~the ESO `European Extremely Large Telescope' (E-ELT) with a 39 m primary mirror, the Giant Magellan Telescope (GMT) with a 24.5m dish, both planned for the southern hemisphere, and the `Thirty Meter Telescope' (TMT) planned for the northern hemisphere, will result in an up to 20-fold increase of signal-to-noise ratio on the spectrum for similar averaging times. Such a gain will yield an immediate improvement of the statistical uncertainty. If currently we measure $\Delta\mu/\mu$ with typical uncertainties of $\sim(5-7) \times10^{-6}$, at a 20-fold increase of SNR a precision level of $(0.3-0.4) \times10^{-6}$ could be reached. This is comparable with the most precise $\Delta\mu/\mu$ constraints obtained from radio measurements of methanol and ammonia at lower redshifts.

Advances in light collection and increase of statistical significance should be accompanied by improvements on the wavelength calibration, the major source of systematic uncertainties. Currently, wavelength calibration is performed by comparison with reference spectra from ThAr lamps. This application is hindered by saturation and asymmetries of calibration lines, but a careful selection of lines warrants a reliable calibration for the present needs and level of accuracy \cite{Murphy2007}. The use of frequency comb lasers for calibration of astronomical spectrographs, now being explored and implemented~\cite{Murphy2007a,Steinmetz2008,Wilken2012}, will provide an ultra-stable, controlled and dense set of reference lines, signifying a major improvement over ThAr calibration. This technical advance will lead to more tightly constraining bounds on $\Delta\mu/\mu$.

The recently encountered long-range wavelength-distortion problems are likely caused by differences in beam pointing between light from the quasar sources and light from the ThAr-emission lamps, rather than from the spectroscopic accuracy of ThAr lines. Differences in pointing between light from distant point sources (quasars) and nearby laser sources may similarly give rise to offsets in calibration. For this reason novel spectrographs will be built for ultimate pressure and temperature control and, most importantly, are designed to be fiber-fed rather than slit-based. This holds for the proposed instruments on the E-ELT and GMT but also for the Echelle SPectrograph for Rocky Exoplanet and Stable Spectroscopic Observations (ESPRESSO) \cite{Pepe2010}, to be mounted and operated on the existing VLT in 2016.  This fiber-fed, cross-dispersed, high-resolution (resolving power $\lambda/\Delta\lambda = 120,000$ and a very high resolution mode at $\lambda/\Delta\lambda = 220,000$) echelle spectrograph will receive light from either one or all four VLT's via a coud\'{e} train optical system. This system will be ideally suited for investigation of high-redshift H$_2$ quasar spectra.
However, an issue with the fiber feed to ESPRESSO and with most future fiber-fed spectrographs mounted on large telescopes is the wavelength coverage. Fiber transmission in the range 3000--3800 \AA\ is low, and the current design window for ESPRESSO only covers 3800--6860 \AA. This implies that a redshift as high as $z=3.1$ is required to observe the entire H$_2$ spectrum until the Lyman cutoff, while at the red end spectral lines at 1140 \AA\ should be captured within the window below 6860 \AA, i.e.~this is for redshifts $z<5$.
These limitations on redshift are rather restrictive in view of potential targets as listed in Table~\ref{Table-quasars}.

Next, one may consider the optimal resolution to be desired for the H$_2$ absorption method while using a spectrograph.
For the existing 8--10m optical telescopes, the designed properties for the spectrograph are entirely matched to the degree to which light from point sources like quasars are blurred out by turbulence and refraction effects in our atmosphere. The angular extent of this `seeing' is typically 0.8--1.0 arcseconds projected on the sky. The entrance slit for spectrographs are matched to this size so that they deliver an instrumental resolution of ${\rm FWHM}\sim6$\,km\,s$^{-1}$, or resolving power $R\sim50,000$. This is sufficient to resolve a single, isolated velocity component in an absorption system with total Doppler parameter of $b=\sqrt{2}\sigma \approx 0.60 \, {\rm FWHM}\sim3$--4\,km\,s$^{-1}$ (where ${\rm FWHM}\equiv 2\sqrt{2\ln 2}\sigma$) from kinetic and turbulent motions.

\begin{figure}[!hb]
\centering
\includegraphics[width=1.0\columnwidth]{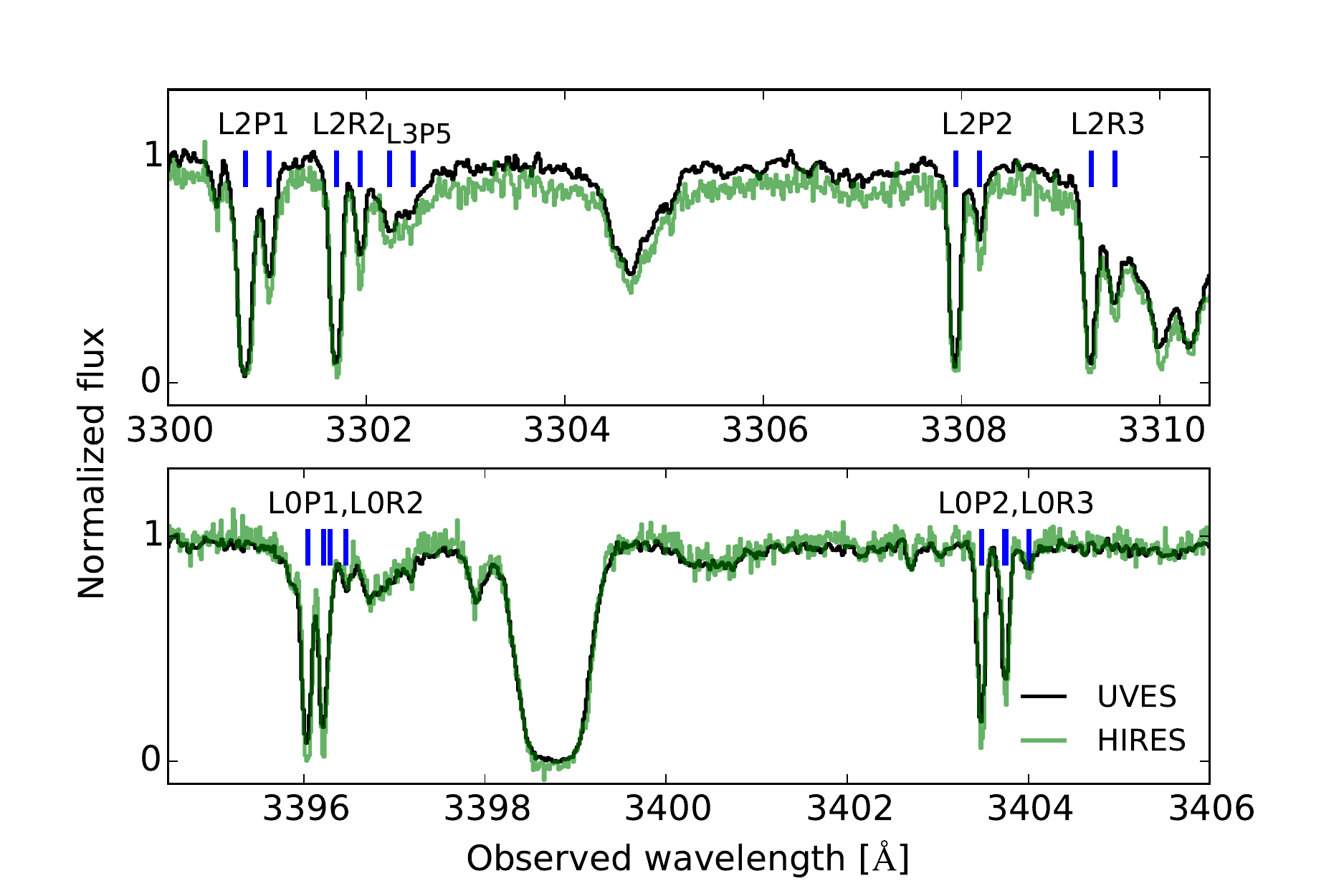}
\caption{Comparison of a part of the spectrum of the absorption system toward J2123$-$005 at $z=2.05$ by the Keck telescope equipped with the HIRES spectrograph, indicated by a grey (green) line  \cite{Malec2010}, recorded at a resolving power of 110,000, and by the Very Large Telescope, equipped with the UVES spectrograph indicated by a dark (black) line \cite{Weerdenburg2011} at a resolving power of 53,000. Note the absence of improved effective resolution in the Keck spectrum, and the higher SNR in the UVES spectrum.}
\label{J2123}
\end{figure}

For a discussion of desired resolution it is illustrative to compare observations of J2123$-$005 at two very different resolution settings. This quasar was observed with the HIRES spectrometer at Keck at a resolution of $R=110,000$ \cite{Malec2010} and with the UVES spectrometer at VLT at $R=53,000$~\cite{Weerdenburg2011}. The extremely high resolution at HIRES could only be achieved due to unrivalled seeing conditions of 0.3 arcseconds. The spectra, shown in Fig.~\ref{J2123}, demonstrate that the factor of two better resolution in the HIRES-Keck spectrum does not effectively produce narrower linewidths.
In the case of J2123$-$005 two very strong and narrow velocity features build the spectrum. However, it is unlikely that each of those features is made up of just a single velocity component. In reality, there may be several components separated by less than $\sim$2\,km\,s$^{-1}$, which are blended together. So, even if each of them only has a Doppler $b$ parameter of $\sim$0.5~km\,s$^{-1}$ or less, they combine to make a flux profile that, even at the very highest resolution, would most likely not be different to that obtained with $R \sim 50,000$. While this may be the case for most H$_2$  absorbers, it is possible that some may have isolated features which truly are made up of just a single, narrow component. Possibly, one of the absorbers HE0027$-$1836, Q0405$-$443 and Q0347$-$383 might be such a case, which could be verified by recording higher resolution spectra of them.
The ESPRESSO spectrograph, and also the PEPSI spectrograph~\cite{Strassmeier2015} with a resolution of $R=270,000$ planned for the Large Binocular Telescope, might be used for testing this hypothesis.
If a quasar system would be found exhibiting narrower line profiles their analysis would definitely yield a more constraining value for $\Delta\mu/\mu$, provided that good signal-to-noise can be obtained and calibration issues can be dealt with.
But in most cases, if not all, a resolving power larger than 50,000 will not by itself lead to significantly improved constraints on $\Delta\mu/\mu$.

\section{Other approaches to probe $\mu$-variation}

The focus of the present review is on the application of the H$_2$ method to observations of quasar spectra with Earth-based telescopes to detect a possible variation of the proton-electron mass ratio on a cosmological time scale. There are a number of methods and applications that are closely connected to this quest. We have selected four of those that will be discussed here.

\subsection{Gamma Ray Bursts}

Besides quasars, the brief and very luminous gamma-ray burst (GRB) afterglows can also be used in searches for extragalactic H$_2$ absorption. GRBs are thought to mark the violent death of (very) massive stars leaving black holes or strongly magnetized neutron stars. Contrary to QSO sight lines, the GRB DLA usually refers to the (star-forming) host galaxy. Only four detections of H$_2$ absorption in GRB-DLAs are known so far, including GRB\,080607 at $z = 3.04$~\cite{Prochaska2009}, GRB\,120815A at $z = 2.36$~\cite{Kruhler2013}, GRB\,120327A at $z = 2.81$~\cite{Delia2014}, and GRB\,121024A at $z = 2.30$~\cite{Friis2014}. While the quality of the current GRB spectra containing H$_2$ is not sufficient to obtain competitive $\Delta\mu/\mu$ constraints, GRBs can in principle give access to more H$_2$ detections at redshifts as high as $z \gtrsim 4$. Another practical advantage of GRB afterglow over QSO spectra is that their power-law spectra are featureless. QSOs become very rare at redshifts larger than 6.

\subsection{Chameleon fields: Environmental dependencies of $\mu$}

There exist theoretical scenarios predicting fundamental constants to depend on environmental conditions, such as surrounding matter density or gravitational fields. The fields that cause these effects go under the name of chameleon theories \cite{Khoury2004}, signifying a distinct origin of varying constants from dilaton field theories \cite{Bekenstein1982,Barrow2002,Sandvik2002}.
The hydrogen absorption method can similarly be applied to probe the spectrum of H$_2$ under conditions of strong gravity as experienced in the photospheres of white dwarfs,
i.e.~the compact electron-degenerate remnant cores of evolved low-mass stars that exhibit
at their surface a gravitational potential some $\sim10^4$ times stronger than that at the Earth's surface. In a few cases H$_2$ molecules are found under such conditions~\cite{Xu2013}. Observations of white dwarf objects in our Galaxy are not subject to cosmological redshift, although there are small effects of gravitational redshift observable in the spectra.

\begin{figure}[!hb]
\centering
\includegraphics[width=1.0\columnwidth]{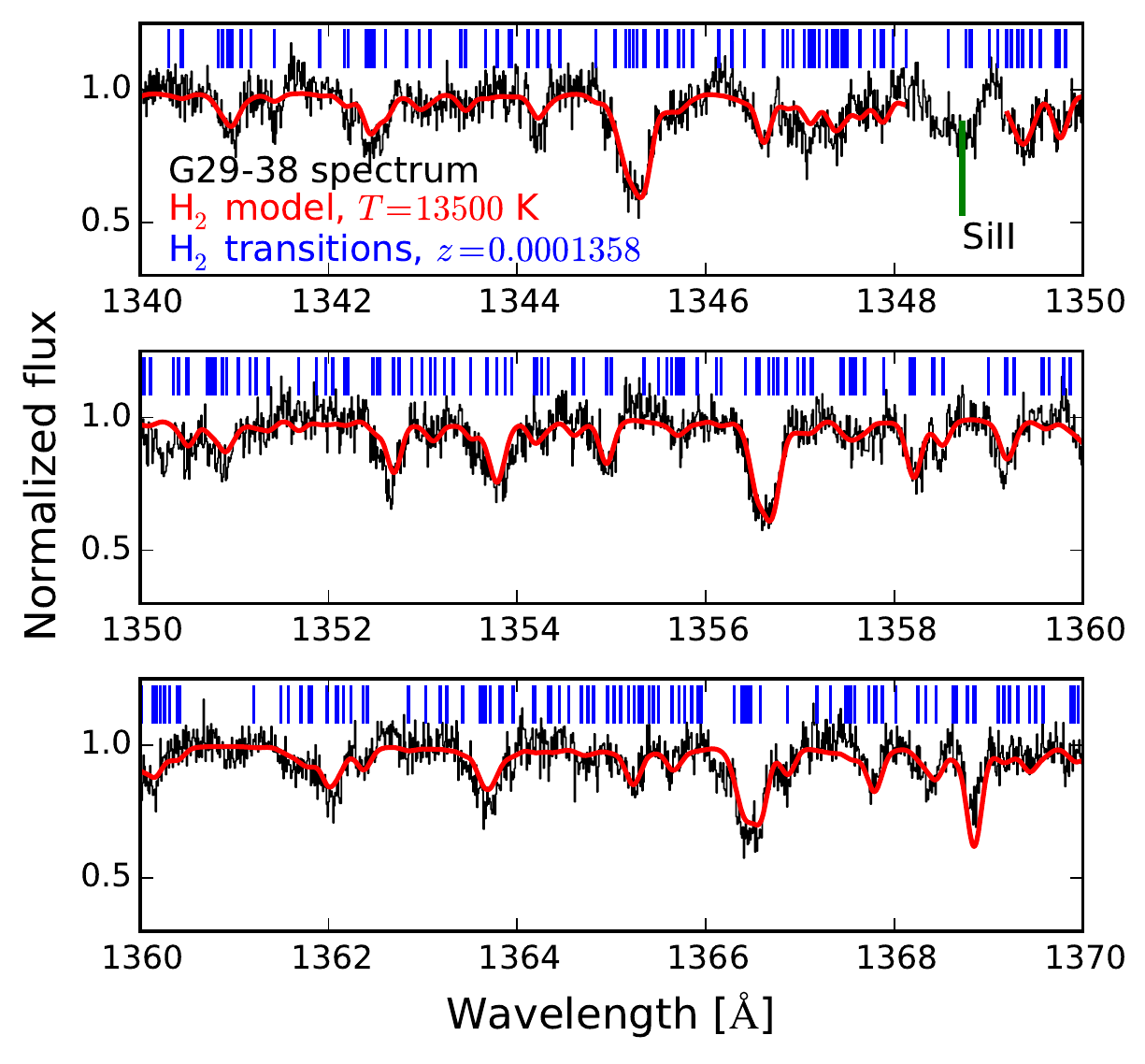}
\caption{Part of the absorption spectrum of the white dwarf G29-38 as obtained with the Cosmic Origins Spectrograph aboard the Hubble Space Telescope in the range \mbox{1340-1370 \AA.} The solid (red) line is the result of a fit with optimized parameters for $T$, $b$, $N$, $z$ and $\Delta\mu/\mu$. The (blue) tick marks represent the positions of H$_2$ absorption lines contributing to the spectrum.}
\label{WD}
\end{figure}

Again Lyman and Werner band lines are the characteristic features, but at the prevailing temperatures of $T \sim 11,000 - 14,000$ K the specific lines contributing to the spectrum are decisively different. It is mainly the $B-X \,(1,v")$ and $B-X \,(0,v")$ Lyman bands for populated vibrational states $v"=3$ and $4$  that contribute to the absorption spectrum. Galactic observation of these bands implies that the absorption wavelengths occur in the vacuum ultraviolet, specifically in the interval $\lambda = 1295 - 1450$\,\AA. Hence, observations from outside the Earth's atmosphere are required, and these were performed with the Cosmic Origins Spectrograph aboard the Hubble Space Telescope~\cite{Xu2013}.
Part of a spectrum of the white dwarf GD29-38 (WD2326$+$049) is shown in Fig.~\ref{WD}.

A $\mu$-variation analysis was performed for the white dwarf spectra based on the same principles as for the quasar observation \cite{Bagdonaite2014b}.
There are, however, remarkable differences mainly related to the difference in excitation temperatures of the absorbing gas. While in quasar absorption systems some 100 lines are observed in the case of the hot photospheres over 1000 lines contribute to the spectrum. A fingerprint spectrum is generated starting from a partition function for a certain temperature $T$:
\begin{equation}
P_{v,J}(T) = \frac{g_n (2J+1) e^{\frac{-E_{v,J}}{kT}}}{\sum\limits_{v=0}^{v_{\mathrm{max}}} \sum\limits_{J=0}^{J_{\mathrm{max}}(v)} g_n (2J+1) e^{\frac{-E_{v,J}}{kT}}}
\label{partition-function}
\end{equation}
where $E_{v,J}$ are excitation temperatures of $(v,J)$ quantum levels, $k$ is the Boltzmann constant and $g_n$ is a nuclear-spin degeneracy factor, for which we adopt a 3:1 ratio for ortho:para levels. Each spectral line, populated by $P_{v,J}$ is then connected to a wavelength $\lambda_i$, an oscillator strength $f_i$, a line damping $\Gamma_i$, while for the entire ensemble a Doppler width $b$ holds. The fingerprint spectrum is then included in a fit to Eq.~(\ref{comp-equation}), introducing a sensitivity coefficient $K_i$ for each line \cite{Salumbides2015}. The parameter $z$ now represents the combined effect of a Doppler velocity of the white dwarf star and a gravitational redshift. A simultaneous fit routine is defined to optimize values of $T$, via Eq.~(\ref{partition-function}), and of $\Delta\mu/\mu$ and $z$ via Eq.~(\ref{comp-equation}) along with  a total H$_2$ column density.


For the two available white dwarf spectra, for GD29-38 and GD133 obtained with COS-HST \cite{Xu2013}, this procedure resulted in bounds on $\mu$-variation  of  $\Delta\mu/\mu_{(\rm{GD29-38})} < (-5.8 \pm 3.7) \times10^{-5}$ and of $\Delta\mu/\mu_{(\rm{GD133})} < (-2.3 \pm 4.7) \times10^{-5}$ ~\cite{Salumbides2015}. The observations provide information on a fundamental constant at a certain value for the gravitational potential, which may be expressed in dimensionless units as:
\begin{equation}
 \phi = \frac{GM}{Rc^2}
\label{gravity}
\end{equation}
with $G$ the Newton-Cavendish gravitational constant, $M$ and $R$ the mass and the radius of the white dwarf, and $c$ the speed of light.
Note that in these dimensionless units the gravitational potential $\phi$ is equal to the gravitational redshift for light traveling out of the white dwarf photosphere $z=GM/Rc^2$. The prevailing conditions for the two white dwarfs at their surface are \mbox{$\phi_{(GD29-38)} = 1.9 \times 10^{-4}$} and \mbox{$\phi_{(GD133)} = 1.2 \times 10^{-4}$}, where the condition at the Earth's surface is \mbox{$\phi_{\rm{Earth}}=0.69 \times 10^{-9}$}. The presently found result may be interpreted as a bound on a dependence of the proton--electron mass ratio $|\Delta\mu/\mu| < 4 \times 10^{-5}$ at field strengths of $\phi > 10^{-4}$. When compared to constraints from Earth-bound experiment with atomic clocks, taking advantage of the ellipticity of the orbit around the sun, the H$_2$ white dwarf studies are less constraining \cite{Blatt2008,Tobar2013}, but bear the advantage of probing regions of much stronger gravity.

\subsection{$\mu$-variation in combination with other fundamental constants}

Besides $\mu$ and $\alpha$, various combinations of dimensionless constants have been probed via spectroscopy of the extragalactic interstellar and/or intergalactic medium. For example, by comparing the ultraviolet (UV) transitions of heavy elements with the hyperfine H\,\textsc{i} transition at a 21-cm rest wavelength allows to extract limits on $x = \alpha^2 g_p/\mu$, where $g_p$ is the gyromagnetic factor of the proton~\cite{Tzanavaris2005}. To make such a comparison, one must take into account the fact that quasars can have frequency-dependent structure, which may result in different sightlines being probed at the UV/optical and radio wavelengths. An application of this particular method led to a $|\Delta x/x|$ constraint at the level of $\lesssim1.7\times10^{-6}$ for redshift $z=3.174$~\cite{Srianand2010}.

Comparison of nearby lying spectral lines of the CO molecule (the 7-6 rotational transition) and the C\,\textsc{i}  (2-1) fine structure line, both observed in emission towards the lensed galaxy HLSJ091928.6$+$514223, constrain the possible variation of another combination of dimensionless constants. Where the CO molecular line exhibits a linear sensitivity to $\mu$-variation, the atomic carbon fine structure splitting exhibits a sensitivity to $\alpha^2$. The combined analysis of the two lines sets a bound on the parameter $F=\alpha^2/\mu$ of
$|\Delta F/F| < 2 \times 10^{-5}$ at a redshift of $z=5.2$ \cite{Levshakov2012}. Similar less-constraining studies on the CO/C\,\textsc{i} combination have been performed as well, even up to redshift $z=6.42$ \cite{Levshakov2008} and over the interval $z=2.3-4.1$ \cite{Curran2011}. Although this method bears the prospect of providing tight constraints on varying constants it relies on the assumption that CO and C\,\textsc{i} are co-spatial.

\subsection{Radio astronomy}

Molecules exhibit many low-frequency modes associated with vibration, rotation and inversion motion. While pure rotational motion exhibits a sensitivity coefficient of $K=-1$ and vibrational motion $K=-1/2$ there are many examples of sensitivity enhancement due to fortuitous degeneracies in the quantum level structure of molecules \cite{Jansen2014,Kozlov2013b}. There are two prominent examples of importance to search for $\mu$-variation via radio astronomy: the ammonia (NH$_3$) and the methanol (CH$_3$OH) molecules.

In NH$_3$, the inversion tunneling of the N-atom through the plane of the H atoms gives rise to transitions that scale with $\mu$ as $E_{\rm{inv}} \sim \mu^{-4.46}$~\cite{Veldhoven2004,Flambaum2007}. The sensitivity coefficient of $K=-4.46$ makes ammonia a two orders of magnitude more sensitive probe for varying $\mu$ than H$_2$. A disadvantage of the ammonia method is that all inversion lines exhibit a similar sensitivity, hence a comparison with different molecules must be made in order to solve for a degeneracy with redshift. Usually, the $K_i=-1$ rotational transitions of other species, as HC$_3$N and HCO$^+$, are picked as the reference transitions. The assumption that the multiple species occupy exactly the same physical location and physical conditions is then a source of systematic uncertainty. Currently, two examples of NH$_3$ absorption are known outside the local universe: at $z=0.69$ towards B0218$+$357 and at $z=0.89$ towards PKS1830$-$211. Measurements of NH$_3$ resulted in 3-$\sigma$ limits of $|\Delta\mu/\mu| < 1.8\times 10^{-6}$ in the former sightline~\cite{Murphy2008a} and $<1.4\times10^{-6}$ in the latter~\cite{Henkel2009}.

Methanol is an even more sensitive probe of varying $\mu$, with $K_i$ sensitivities ranging from $-33$ to $+19$~\cite{Jansen2011,Levshakov2011}. Besides purely rotational transitions, the microwave spectrum of CH$_3$OH includes rotation-tunneling transitions arising via the hindered rotation of the OH group, thus inducing the greatly enhanced sensitivity to $\mu$. In fact, the most stringent current $|\Delta\mu/\mu|$ constraints at the level of $<1.1\times 10^{-7}$ (1-$\sigma$) are derived from a combination of these CH$_3$OH absorption lines at redshift $z=0.89$ towards PKS1830$-$211~\cite{Bagdonaite2013a,Bagdonaite2013b,Ellingsen2012,Kanekar2015}. As opposed to the NH$_3$ method, where a comparison with other molecular species is often used, methanol offers an advantage of a test based on a single species. Unfortunately methanol has only been detected in the single lensed galaxy PKS1830$-$211, where it is subject to a number of
disturbing phenomena such as time variability of the background blazar and chromatic substructure \cite{Muller2008,Bagdonaite2013b}.

As of now only two lensed galaxies have been found with a sufficient radiation level and column density to detect NH$_3$ and CH$_3$OH.
Apart from PKS1830$-$211 and B0218$+$357, there exist three further intermediate redshift radio sources where molecules have been detected, i.e. PKS1413$+$357 with absorption at $z=0.247$~\cite{Wiklind1994}, B1504$+$377 with absorption at $z=0.672$~\cite{Wiklind1996b}, and PMN J0134$-$0931 with absorption at $z=0.765$~\cite{Kanekar2005}. PMN J0134$-$0931 is also a gravitationally lensed object. The high detection-rate of gravitationally lensed objects is a selection bias because lensing amplifies the flux of the background quasar. It is expected that with the recently inaugurated and planned radio-astronomical facilities, such as the Atacama Large Millimeter/submillimeter Array (ALMA; frequency range 30 GHz - 1 THz) or the Square Kilometre Array (SKA; frequency range up to 30 GHz in the final stage), more detections of extragalactic molecular absorption will be found in the foreseeable future. The enhanced sensitivity of the new cm- and mm-wave telescopes will allow finding weaker radio sources, and the broad bandwidths of the new spectrometers will permit sensitive studies of varying constants at high redshift.

\section{Conclusion}

The H$_2$ quasar absorption method is demonstrated to be a viable method to search for putative variations of the proton--electron mass ratio $\mu$, in particular for redshift ranges $z>2$. The method is based on a comparison between laboratory and astronomical absorption spectra of Lyman and Werner lines in molecular hydrogen. The wavelengths of the laboratory lines are determined at the accuracy level of $\Delta\lambda/\lambda < 10^{-7}$ and may therefore be considered exact for the purpose of these comparisons, i.e.~not contributing to the uncertainty on $\Delta\mu/\mu$. Sensitivity coefficients, representing the differential wavelength offset of the spectral lines induced by a shift in $\mu$, are calculated to an accuracy of $< 1\%$ so that they also do not cause a limitation to a determination of $\Delta\mu/\mu$. By now ten H$_2$ absorption systems have been analyzed in detail. They deliver an averaged quantitative result of $\Delta\mu/\mu = (3.1 \pm 1.6)\times 10^{-6}$ in the redshift  interval $z=2.0-4.2$. Part of this positive value at the 1.9-$\sigma$ level might be explained by some not yet fully included long-range wavelength distortions in the astronomical calibrations. Hence we conservatively interpret this result as a null-result of $|\Delta\mu/\mu| < 5 \times 10^{-6}$ (3-$\sigma$), meaning that the proton--electron mass ratio has changed by less than 0.0005\% for look-back times of 10--12.4 billion years into cosmic history.

Attempts to interpret the results from the ten H$_2$ absorbers in terms of a spatial variation of $\mu$ along the dipole axis ($RA = 17.5$ hrs and dec =$ -58$ degrees), found from analyzing 300 metal absorption systems for $\alpha$ at high redshift, did not yield a reliable result because of the small sample available and the coincidental alignment of these systems in a narrow band across the sky.

Prospects of improving results by the H$_2$ method are identified. The upcoming opportunities of larger class telescopes (E-ELT, GMT and TMT) will deliver increased photon collection, where signal-to-noise improvements will straightforwardly return tighter bounds on $\Delta\mu/\mu$. Most important is, however, improvement of calibration methods for the astronomical data. The combination of frequency-comb calibration with a fiber-fed spectrometer, as already planned for the ESO-VLT (ESPRESSO spectrometer), would resolve the existing calibration issues. Included coverage of the short-wavelength range [3000-3800] \AA\ would be very much desired since many of the identified high-quality H$_2$ absorption systems (in the redshift range $z=2-3$) have their relevant features in this window.

As a final note we state that even incremental improvements setting boundaries on drifting fundamental constants are worthwhile to pursue, given the importance of this endeavor into the nature of physical law: Is it constant or not?

\section*{Acknowledgements}

The Netherlands Foundation for Fundamental Research of Matter (FOM) is acknowledged for financial support through its program `Broken Mirrors \& Drifting Constants'. WU acknowledges support from the Templeton Foundation (New Frontiers in Astronomy and Cosmology program). This project has received funding from the European Research Council (ERC) under the European
Union's Horizon 2020 research and innovation programme (grant agreement No 670168).
MTM thanks the Australian Research Council for \textsl{Discovery Project} grant DP110100866 which supported this work.


\end{document}